\documentclass[final]{siamart0516}

\usepackage{lipsum}
\usepackage{amsfonts}
\usepackage{graphicx}
\usepackage{epstopdf}
\usepackage{algorithmic}
\ifpdf
  \DeclareGraphicsExtensions{.eps,.pdf,.png,.jpg}
\else
  \DeclareGraphicsExtensions{.eps}
\fi

\newcommand{\TheTitle}{A tetrahedral space-filling curve for non-conforming adaptive meshes}
\newcommand{\TheShortTitle}{A tetrahedral space-filling curve}
\newcommand{\TheAuthors}{C, Burstedde, J. Holke}

\headers{\TheShortTitle}{\TheAuthors}

\title{{\TheTitle} 
}
\author{
  Carsten Burstedde\thanks{Institut  f{\"u}r Numerische Simulation (INS) %
    and Hausdorff Center for Mathematics (HCM), %
    Rheinische Friedrich-Wilhelms-Universit{\"a}t Bonn, Germany.}
  \and
  Johannes Holke%
  \footnotemark[1]\hspace{1ex}%
  \thanks{Corresponding author (\email{holke@ins.uni-bonn.de})}
}

\usepackage{amsopn}

\ifpdf
\hypersetup{
  pdftitle={\TheTitle},
  pdfauthor={\TheAuthors}
}
\fi

\usepackage{cite} 
\usepackage{mathrsfs} 
\usepackage[all,2cell]{xy} \UseAllTwocells \SelectTips{eu}{10} 
\usepackage{multirow}	  
\usepackage[usenames,dvipsnames]{xcolor} 	
\usepackage{enumerate}
\usepackage[ruled,linesnumbered,algosection,vlined]{algorithm2e} 
\usepackage{sidecap}	
\usepackage{hhline}	  

\newcommand{\set}[1]{\left\lbrace\, #1 \,\right\rbrace}
\newcommand{\IL}{{\mathbb{L}}}
\newcommand{\IR}{{\mathbb{R}}}
\newcommand{\IZ}{{\mathbb{Z}}}
\newcommand{\IN}{{\mathbb{N}}}
\newcommand{\abst}[1]{\, #1 \,} 
\newcommand{\mytabvspace}{\vphantom{${X^X}^X$}} 

\newcommand{\dtype}[1]{{#1}\xspace}
\newcommand{\Int}{\dtype{int}}
\newcommand{\Tet}{\texttt{Tet}}
\newcommand{\aTet}{{\Tet}} 
\newcommand{\Comment}[1]{\tcc*[r]{#1}}   
\newcommand{\IfComment}[1]{\tcc*[f]{#1}} 
\newcommand{\algomod}{\abst{\scalebox{0.9}{\%}}} 
\newcommand{\algoand}{\textbf{ and }} 
\newcommand{\algor}{\textbf{ or }} 
\newcommand{\bitwand}{\;\&\;} 
\newcommand{\aorgl}{\abst{|\hspace{-0.5ex}=}} 
\newcommand{\cfbox}[2]{
    \colorlet{currentcolor}{.}%
    {\color{#1}%
    \fbox{\color{currentcolor}#2}}%
  }
\newcommand{\e}{\text{e}} 

\DeclareMathOperator{\inter}{\dot\perp} 
\newcommand{\ainter}{\abst{\inter}}
\DeclareMathOperator{\type}{type}
\DeclareMathOperator{\cid}{\textrm{cube-id}}

\theoremstyle{plain}
\newtheorem{property}[theorem]{Property}
\newtheorem{remark}[theorem]{Remark}

\graphicspath{{../../timings/adapt/}}

\begin{document}

\maketitle

\begin{abstract}
We introduce a space-filling curve for triangular and tetrahedral
red-refinement that can be computed using bitwise interleaving operations similar
to the well-known Z-order or Morton curve for cubical meshes.
To store sufficient information for random access, we define a low-memory
encoding using 10 bytes per triangle and 14 bytes per tetrahedron.
We present algorithms that compute the parent, children, and face-neighbors of
a mesh element in constant time, as well as the next and previous element in
the space-filling curve and whether a given element is on the boundary of the
root simplex or not.
Our presentation concludes with a scalability demonstration that creates and
adapts selected meshes on a large distributed-memory system.
\end{abstract}

\begin{keywords}
  Forest of octrees, parallel adaptive mesh refinement, Morton code,
  high per\-for\-mance computing,
  nonconforming simplicial mesh,
  space-filling curve
\end{keywords}

\begin{AMS}
  65M50, 
  68W10, 
  65Y05, 
  65D18  
\end{AMS}

\section{Introduction}

Conforming adaptive mesh refinement for simplicial (triangular and tetrahedral)
meshes is one of the most successful concepts in numerical mathematics and
computational science and engineering; see, e.g., \cite{BabuskaRheinboldt78,Dorfler96,SteinmanMilnerNorleyEtAl03}.
Simplices provide high flexibility in meshing to arbitrary domain geometries\cite{Shewchuk96, Si06}
and can be mapped
to a reference simplex
using an elementwise constant Jacobian in most cases, which
allows for an efficient numerical implementation.
Discretization and integration methods of various kinds are available for both
low and high orders of accuracy.

Large-scale scientific computing requires fast
and scalable algorithms for (1) adaptive refinement and coarsening (AMR) as well as
(2) parallel partitioning.
One class of methods for AMR exploits the properties of Delaunay triangulations
\cite{GoliasDutton97,Shewchuk97d,HjelleDaehlen06}, while partitioning of
unstructured meshes is often
approached by formulating the mesh topology as a graph.
These triangulations are usually conforming; that is, elements intersect only
along whole faces and edges.
Common application codes such as FEniCS \cite{LoggMardalWells12}, PLUM
\cite{OlikerBiswas98,OlikerBiswasGabow00}, OpenFOAM \cite{OpenCFD07}, or
MOAB from the SIGMA toolkit \cite{TautgesMeyersMerkleyEtAl04} delegate
graph partitioning to third-party software like ParMETIS \cite{KarypisKumar98} or Scotch \cite{ChevalierPellegrini08}.
Graph-based algorithms have been advanced to target millions of processes and
billions of elements
\cite{CatalyurekBomanDevineEtAl07,RasquinSmithChitaleEtAl14,SmithRasquinIbanezEtAl15}.
Still, increasing the scalability and decreasing the absolute runtime and
memory demands of distributed implementations
remains a challenge,
and the lack of an obvious parent-child structure in many unstructured meshing
approaches prevents certain use cases.

Nonconforming AMR in combination with recursive refinement
makes refinement and coarsening nearly trivial operations.
The additional mathematical logic to enable hanging faces and edges is
well understood for both continuous and discontinuous discretizations
\cite{RheinboldtMesztenyi80,FischerKruseLoth02,
KoprivaWoodruffHussaini02,
AkcelikBielakBirosEtAl03}.
It is local to the loop over the finite elements or volumes and transparent to
most of the numerical pipeline, thus offering the possibility to
extend existing conforming codes.
The resolution may be as coarse as any chosen root
mesh (a mesh which is not intended for further coarsening),
which poses only a slight limit to geometric flexibility.

The challenge of efficient partitioning
of meshes may be addressed using space-filling curves (SFC).
Instead of working on the NP-hard graph partitioning problem \cite{BuiJones92},
SFCs are used to approximately solve the partitioning problem in
linear runtime \cite{Bader12,Zumbusch02}.
Common SFCs establish an equivalence between the adaptive mesh and a tree,
where the mesh elements correspond one-to-one to the leaves of the tree.
They also define a total order among the leaves that can be used to store
application data linearly in the same order as the elements
\cite{BursteddeGhattasStadlerEtAl08}.
The original concept is specified for a cubic domain and can be traced back to
over a hundred years ago \cite{Peano90,Hilbert91} .
The Sierpinski curve \cite{Sierpinski12} and generalized versions of it are the
most common 
SFCs on simplicial meshes.
These were described by B{\"a}nsch \cite{Bansch91}, Kossaczk\`y
\cite{Kossaczk`y94}, and others; see e.g.\ \cite{Sagan94,Bader12}.
The computation of parents and children, for example, usually involves a loop
over the refinement levels, equivalent to following the path between the tree
root and a leaf.

When using hypercubes as mesh primitives, the logic of common SFCs can be
formulated with remarkable simplicity due
to the local tensor product structure \cite{Peano90, StewartEdwards04,
TuOHallaronGhattas05, SundarSampathBiros08, Zumbusch02}.
It is also well-suited to compute topological entities like face-neighbors,
children, and parents of given mesh elements.
The Morton curve \cite{Morton66}, in particular, allows one to design
algorithms for these tasks (see, e.g., %
\cite{BursteddeWilcoxGhattas11})
whose runtime is level-independent, i.e., does not depend on the depth of the
refinement tree.
Moreover, SFCs are memory efficient, since for a mesh element only the
coordinates of one anchor node plus its refinement level have to be stored.
Today, this principle has found its way into widely used software libraries
\cite{BangerthHartmannKanschat07,BangerthBursteddeHeisterEtAl11}.
Memory usage can be further optimized by incremental encoding along the SFC
\cite{BungartzMehlWeinzierl06,WeinzierlMehl11,Bader12}.

Using SFCs on hexahedral meshes is exceptionally fast and scalable
\cite{RahimianLashukVeerapaneniEtAl10, BursteddeGhattasGurnisEtAl10,
IsaacBursteddeWilcoxEtAl15}.
This fact has not only been exploited in writing simulation codes using
hexahedral meshes, but also by approaches that use the hexahedral SFC as an
instrument to
partition simplices, mapping them into a surrounding cube
\cite{AlauzetLoseille09}.
However, hexahedral AMR is more limited geometrically than simplicial AMR.
One indication for this is the (lack of) availability of (open-source) mesh
generators that operate exclusively on cubes.
Furthermore, assuming an existing simplicial code, rewriting it in terms of
cubical meshes will be out of the question if the code is of sufficient size,
complexity, or maturity.
For these reasons,
in this paper, we attempt to establish a new SFC for
triangles and tetrahedra that has many of the favorable properties known for
hexahedra.

Our starting point is to divide the simplices in a refined mesh into two 
(two dimensions, 2D),
respectively six (three dimensions, 3D), different types
and selecting for each type a specific reordering for Bey's red-refinement \cite{Bey92}.
This type and the coordinates of one vertex
serve as a unique identifier, the
Tet-id, of the simplex in question.
In particular, we do not require storing the type of all parent simplices to
the root, as one might naively imagine.
We then propose a Morton-like coordinate mapping that can be computed from the
Tet-id and gives rise to an SFC.
Based on this logic, we develop constant-time algorithms
(that is, not requiring a loop over the refinement levels) to
compute the tetrahedral-Morton identifiers of any parent, child, face-neighbor,
and SFC-successor/predecessor and to decide whether for two given elements one
is an ancestor of the other.
We conclude with
scalability tests of mesh creation and adaptation with over $8.5%
\times 10^{11}$ tetrahedral mesh elements on up to 131,072 cores of the
JUQUEEN supercomputer and 786,432 cores of MIRA.

\section{Mesh refinement on simplices}
\label{sec:BeyRef}

Our aim is to define and examine a new SFC for triangles and
tetrahedra by adding ordering prescriptions to the nonconforming
Bey-refinement (also called red-refinement) \cite{Bey92,Bey00,Zhang95}.
We briefly restate the red-refinement in this section and contrast it with the
well-known conforming (or red/green) refinement.

We refer to triangles and tetrahedra as $d$-simplices, where $d\in\set{2,3}$
specifies the dimension.
It is sometimes convenient to drop $d$ from this notation.
A $d$-simplex $T\subseteq \IR^d$ is uniquely determined by its $d+1$
affine-independent corner nodes $\vec x_0,\dots,\vec x_{d} \in \IR^d$.
Their order is significant, and therefore we write
\begin{subequations}
\begin{align}
  T& = [\vec x_0,\vec x_1,\vec x_2] \hphantom{,x_3} \quad\textrm{in 2D},\\
  T& = [\vec x_0,\vec x_1,\vec x_2,\vec x_3] \quad\textrm{in 3D}.
\end{align}
\end{subequations}
We define $\vec x_0$ as the \emph{anchor node} of $T$.
By $\vec x_{ij}$ we denote the midpoint between $\vec x_i$ and $\vec x_j$,
thus $\vec x_{ij}=\frac{1}{2}(\vec x_i+\vec x_j)$.

\begin{figure}
\center
\begin{minipage}{0.48\textwidth}
   \def\svgwidth{40ex}
   \includegraphics{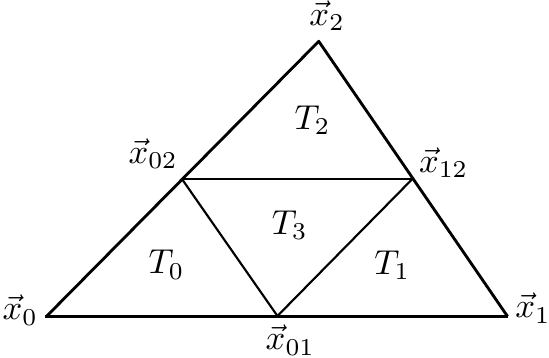}
\end{minipage}
\begin{minipage}{0.48\textwidth}
   \def\svgwidth{40ex}
   \includegraphics{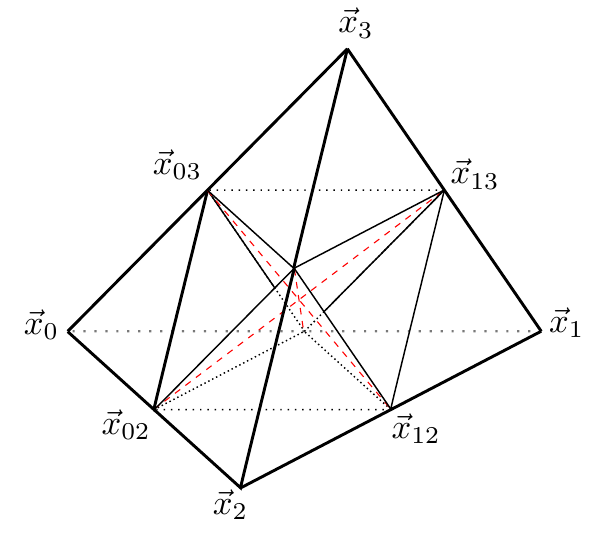}
\end{minipage}
   \caption{%
   Left: The refinement scheme for triangles in two dimensions. A triangle $T=[\vec x_0,\vec x_1,\vec x_2]\subset\IR^2$ is refined by
   dividing each face at the midpoints $\vec x_{ij}$. We  obtain four smaller triangles, all similar to $T$.
   Right: The situation in three dimensions. If we divide the edges of the tetrahedron $T=[\vec x_0,\vec x_1,\vec x_2,\vec x_3]\subset\IR^3$ in half,
   we get four smaller tetrahedra (similar to $T$) and
   an inner octahedron.
   By dividing the octahedron along any of its three diagonals (shown
   dashed) we finally end up with
   partitioning $T$ into eight smaller tetrahedra, all having the same volume.
   The refinement rule of Bey is obtained by always choosing the diagonal from $\vec x_{02}$ to
   $\vec x_{13}$ and numbering the corners of the children according to \eqref{eq:childnumbers}.}
   \label{fig:cutoff}
\end{figure}%

\subsection{Bey's refinement rule}

Bey's rule is a prescription for subdividing a simplex.
It is one instance of the so-called red-refinement, where all faces of a
simplex are subdivided simultaneously.
\begin{definition}
 Given a $d$-simplex $T=[\vec x_0,\dots,\vec x_d]\subset\IR^d$, the refinement
 rule of Bey consists of cutting off four subsimplices at the corners
 (as in Figure \ref{fig:cutoff}).
 In 3D the remaining octahedron is then divided along the diagonal from $\vec
 x_{02}$ to $\vec x_{13}$.
 Bey numbers the $2^d$ resulting subsimplices as follows.
 \begin{subequations}
\label{eq:childnumbers}
 \begin{equation}
 \label{eq:childnumbers_2d}
 2D:\quad
 \begin{array}{cccccl}
  T_0&:=&[\vec x_0,\vec x_{01},\vec x_{02}], &  T_1&:=&[\vec x_{01},\vec x_{1},\vec x_{12}],\\
  T_2&:=&[\vec x_{02},\vec x_{12},\vec x_{2}], & T_3&:=&[\vec x_{01},\vec x_{02},\vec x_{12}],
  \end{array}
 \end{equation}
 \begin{equation}
\label{eq:childnumbers3d}
 3D:\quad
 \begin{array}{cccccc}
 T_0 &:=& [\vec x_0,\vec x_{01},\vec x_{02},\vec x_{03}],   & T_4 &:=& [\vec x_{01},\vec x_{02},\vec x_{03},\vec x_{13}],\\
 T_1 &:=& [\vec x_{01},\vec x_{1},\vec x_{12},\vec x_{13}], & T_5 &:=& [\vec x_{01},\vec x_{02},\vec x_{12},\vec x_{13}],\\
 T_2 &:=& [\vec x_{02},\vec x_{12},\vec x_{2},\vec x_{23}], & T_6 &:=& [\vec x_{02},\vec x_{03},\vec x_{13},\vec x_{23}],\\
 T_3 &:=& [\vec x_{03},\vec x_{13},\vec x_{23},\vec x_{3}], & T_7 &:=& [\vec x_{02},\vec x_{12},\vec x_{13},\vec x_{23}].
\end{array}
 \end{equation}
 \end{subequations}
\end{definition}
\begin{definition}
In this paper, a \emph{refinement} of a $d$-simplex $S$ denotes a set $\mathscr S$
of $d$-dimensional subsimplices of $S$ that can be constructed from 
$S$ via successive refinement, where only the finest simplices belong to the 
actual refinement.
Thus all refinements can be obtained applying the following rules:
\begin{itemize}
 \item $\mathscr{S}=\set{S}$ is a refinement of $S$.
 \item If $\mathscr{S'}$ is a refinement of $S$, then
    $\mathscr{S}=\left(\mathscr{S'}\backslash \set T\right) \cup \set{T_0,\dots,T_{2^d-1}}$
    is a refinement for every $T\in\mathscr S'$.
\end{itemize}
We explicitly allow nonuniform refinements and thus nonconforming faces and edges.
\end{definition}
\begin{definition}
\label{def:childrenancestor}
The $T_i$ from \eqref{eq:childnumbers} are called the \emph{children} of $T$,
and $T$ is called the \emph{parent} of the $T_i$, written $T=P(T_i)$.
Therefore, we also call the $T_i$ \emph{siblings} of each other.
If a $d$-simplex $T$ belongs to a refinement of
another $d$-simplex $S$, then $T$ is a \emph{descendant} of $S$, and $S$ is an \emph{ancestor} of $T$.
The number $\ell$ of refining steps needed to obtain $T$ from $S$ is unique
and called the \emph{level} of $T$ (with respect to $S$); we write $\ell=\ell(T)$.
Usually $S$ is clear from the context, and therefore we %
omit it in the notation.
By definition, $T$ is an ancestor and descendant of itself.
\end{definition}

\begin{figure}
  \begin{minipage}{0.60\textwidth}
   \includegraphics[width=0.8\textwidth]{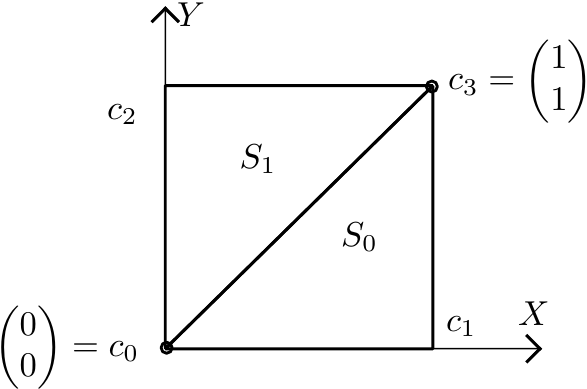}

   \includegraphics[width=0.8\textwidth]{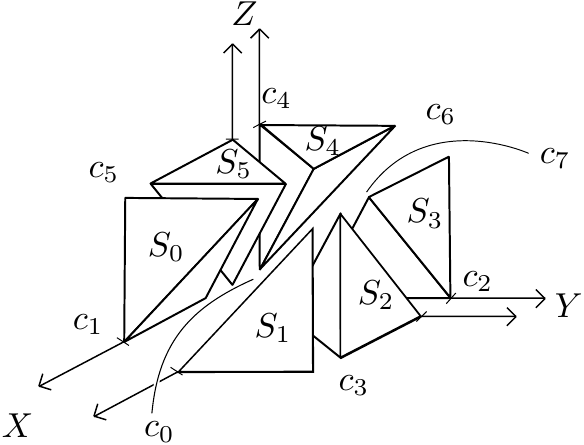}
 \end{minipage}
  \raisebox{7ex}{%
  \begin{minipage}{0.39\textwidth}
   \flushright
  \begin{subequations}
  \label{eq:coordsofSb}
    \begin{equation}
    \label{eq:coordsofSb_2d}
\begin{array}{r|cccc}
2\textrm D& \vec x_0 & \vec x_1 & \vec x_2 \\[0.5ex]\hline
 \vphantom{{X^X}^X}S_0& c_0 & c_1 & c_3 \\[0.2ex]
 S_1& c_0 & c_2 & c_3
\end{array}
\end{equation}\\[14ex]
  \begin{equation}
  \label{eq:coordsofSb_3d}
\begin{array}{r|cccc}
 3\textrm D& \vec x_0 & \vec x_1 & \vec x_2 & \vec x_3 \\[0.5ex]\hline
 \vphantom{{X^X}^X}S_0& c_0 & c_1 & c_5 & c_7 \\[0.2ex]
 S_1& c_0 & c_1 & c_3 & c_7 \\[0.2ex]
 S_2& c_0 & c_2 & c_3 & c_7 \\[0.2ex]
 S_3& c_0 & c_2 & c_6 & c_7 \\[0.2ex]
 S_4& c_0 & c_4 & c_6 & c_7 \\[0.2ex]
 S_5& c_0 & c_4 & c_5 & c_7
\end{array}
\end{equation}
\end{subequations}
 \end{minipage}}
 \caption{The basic triangle (2D) and tetrahedra types (3D) obtained by dividing $[0,1]^d$ into simplices of varying types, denoted by a subscript.
        Top left: The unit square can be divided into two triangles sharing the
edge from $(0,0)^T$ to $(1,1)^T$. We denote these triangles by $S_0$ and $S_1$.
The four corners of the square are numbered $c_0,\ldots,c_3$ in $yx$-order.
Top right: The corner nodes of $S_0$ and $S_1$ in terms of the square corners.
Bottom left (exploded view): In three dimensions the unit cube can be divided
into six tetrahedra, all sharing the edge from the origin to $(1,1,1)^T$.  We
denote these tetrahedra by $S_0,\dots,S_5$.  The eight corners of the cube are
numbered $c_0,\ldots,c_7$ in $zyx$-order (redrawn and modified with permission
\cite{Bey92}).
        Bottom right: The corner nodes of the six tetrahedra $S_0,\dots,S_5$ in
terms of the cube corners.} 
\label{fig:sechstetra}
\end{figure}

\begin{figure}
\xymatrix@C=16ex@R=15ex{
*+<4pt,4pt>{
   \includegraphics[width=15ex]{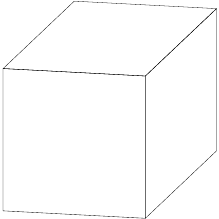}
   }
   \ar[r]^-{\textrm{refining the cube}}\ar[d]|{\textrm{triangulating the cube}}
   &
   *+<4pt,4pt>{
   \includegraphics[width=15ex]{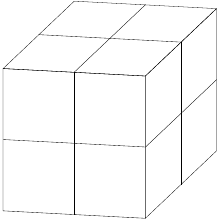}
   }
   \ar[d]|{\textrm{triangulating each cube}}
   \ar[r]^-{\textrm{refining each cube}}
   &
   *+<4pt,4pt>{
   \includegraphics[width=15ex]{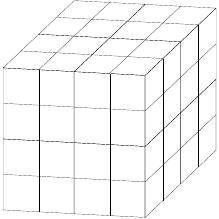}
   }
   \ar[d]|{\textrm{triangulating each cube}}
   \\
  *+<4pt,4pt>{
   \includegraphics[width=15ex]{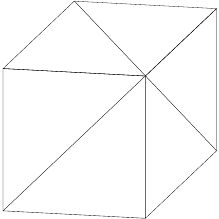}
   }
   \ar[r]^-{\textrm{refining}}_-{\textrm{each tetrahedron}} &
  *+<4pt,4pt>{
  \includegraphics[width=15ex]{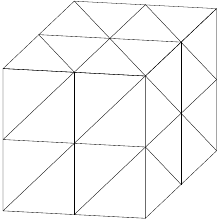}
   }
   \ar[r]^-{\textrm{refining}}_-{\textrm{each tetrahedron}} &
    *+<4pt,4pt>{
    \includegraphics[width=15ex]{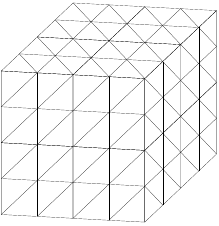}
    }
}
\caption{Triangulating a cube according to Figure \ref{fig:sechstetra} and then refining the tetrahedra via Bey's refinement rule
     results in the same mesh as first refining the cube into eight subcubes and afterward triangulating
     each of these cubes.
     Each occurring tetrahedron is uniquely determined by the subcube it lies in
plus its type.  The same situation can be observed in 2D if we
restrict our view to one side of the cube.}
\label{fig:refineddiagram}
\end{figure}

Consider the six tetrahedra $S_0,\dots,S_5\abst{\subset} \IR^3$
displayed in Figure \ref{fig:sechstetra}.
These tetrahedra form a triangulation of the unit cube.
The results and algorithms in this paper rely on the following property
\cite{Bey92}.
\begin{property}
\label{property:commref}
Refining the six tetrahedra from the triangulation of the unit cube simultaneously to level $\ell$ results in the same
mesh as first refining the unit cube to level $\ell$ and then
triangulating each smaller cube with the six tetrahedra $S_0,\dots,S_5$, scaled
by a factor of $2^{-\ell}$ (see Figure \ref{fig:refineddiagram}).
The same behavior can be observed in 2D when the unit square is
divided into two triangles.
\end{property}

\subsection{Removal of hanging nodes using red/green refinement}

It is worth noting that, although the methods and algorithms presented in this
paper apply to red-refined meshes with hanging nodes, it is possible to augment
them to create meshes without hanging nodes.
For this we may use red/green or red/green/blue refinement methods
\cite{AndrewShermanWeiser83,Carstensen04a}.

\begin{figure}
\centering
  \begin{minipage}{0.40\textwidth}
\centering
  \includegraphics[width=0.8\textwidth]{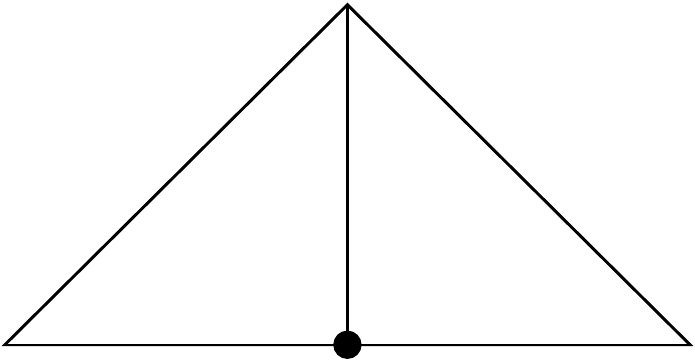}
\end{minipage}
 \begin{minipage}{0.40\textwidth}
  \centering
  \includegraphics[width=0.8\textwidth]{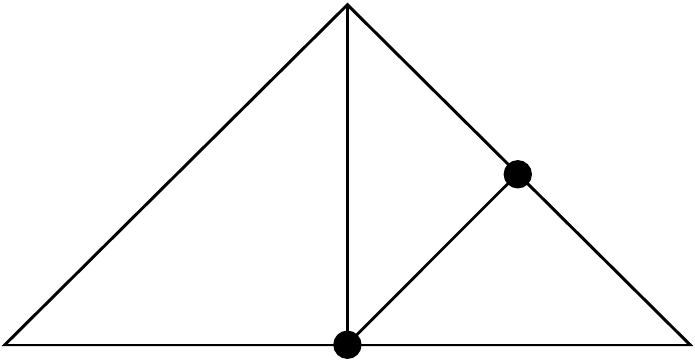}
\end{minipage}
\caption{To resolve hanging nodes we can execute an additional step of green- or
    blue-refinement after the last refinement step \cite{AndrewShermanWeiser83,
    Carstensen04a}.
    Here we show the 2D refinement rules.
    Left: green (1 hanging node).
    Right: blue (2 hanging nodes).
    If a triangle has 3 hanging nodes it is red-refined.}%
    \label{fig:greenblueref}
\end{figure}

After the red-refinement step we may add an additional and possibly nonlocal
refinement operation that ensures a maximum level difference of 1 between
neighboring simplices.
Such an operation is also called 2:1 balance
\cite{IsaacBursteddeGhattas12,SundarSampathBiros08,TuOHallaronGhattas05};
its full description, however, would exceed the scope of the present paper.
Hanging nodes are then resolved by bisecting
those simplices with hanging nodes (green/blue refinement)
\cite[section 12.1.3]{Bader12}.
The 2D case is shown in Figure \ref{fig:greenblueref}. 

If one of the newly created simplices shall be further refined, the bisection
is reversed, the original simplex is red-refined, and the balancing and
green-refinement is repeated.
This may void the nesting property of certain discrete function spaces, yet
applications may still prefer this approach over the manual implementation of
hanging node constraints.

\section{A tetrahedral Morton index}
The Morton index or Z-order for a cube in a hexahedral mesh is computed by bitwise interleaving the coordinates of
the anchor node of the cube \cite{Morton66}.
In this section we present an index for $d$-simplices that also uses the
bitwise interleaving approach, the tetrahedral Morton index (TM-index).
To define the TM-index we look at refinements of a reference simplex, which we
discuss in Section~\ref{sec:reference} below.
For each $d$-simplex in a refinement of the reference simplex we define a
unique identifier, the so-called Tet-id, which serves as the input to compute
the TM-index and for all algorithms related to it.
This Tet-id consists of the coordinates of the anchor node of the considered simplex
plus one additional number, the type of the simplex.
We define the Tet-id and type in Section \ref{sec:type_id}.
We then define the TM-index in Section \ref{sec:TMindex} and in the following
subsections.
We show that it possesses
locality properties similar to those of the Morton index.
One novel aspect of this construction lies in logically including the types of
the simplex and all its parents in the interleaving, while only using the type
of the simplex itself in the algorithms.

\subsection{The reference simplex}
\label{sec:reference}

Throughout the rest of this paper, let $\mathcal{L}$ be a given maximal refinement level.
Instead of the unit cube $[0,1]^d$, we consider the scaled cube $[0,2^\mathcal{L}]^d$, ensuring that all node coordinates
in a refinement up to level $\mathcal L$ are integers.
Suppose we are given some $d$-simplex $T\subset\IR^d$ together with a refinement $\mathscr S$ of $T$.
By mapping $T$ affine-linearly to $2^\mathcal{L} S_0$ the refinement $\mathscr S$ is mapped
to a refinement $\mathscr S'$ of $2^\mathcal{L} S_0$.
Therefore, to examine SFCs on refinements of $T$, it suffices to examine SFCs
on $2^\mathcal{L} S_0$.
Thus, we only consider refinements of the $d$-simplex $T^0_d:=2^\mathcal{L} S_0$.
Let $\mathcal{T}_d$  be the set of all possible descendants of this $d$-simplex with level smaller than or
equal to $\mathcal{L}$; thus
\begin{equation}
 \mathcal{T}_d\abst{=} \set{T \abst{|} T \textnormal{ is a descendant of } T^0_d
 \textnormal{ with } 0\leq \ell(T)\leq\mathcal{L}}.
\end{equation}
Any refinement (up to level $\mathcal L$) of $T^0_d$ is a subset of $\mathcal T_d$, and for each $T\in\mathcal T_d$ there exists at least one
refinement $\mathscr T$ of $T^0_d$ with $T\in\mathscr T$.
In this context, we refer to $T^0_d$ as the \emph{root simplex}.
Furthermore, let $\IL^d$ denote the set of all possible anchor node coordinates of $d$-simplices in $\mathcal{T}_d$, thus
\begin{equation}
 \begin{array}{ccl}
  \IL^2&=&\set{[0,2^\mathcal{L})^2\cap \IZ^2 \abst | y\leq x},\\
  \IL^3&=&\set{[0,2^\mathcal{L})^3\cap \IZ^3 \abst | y\leq z \leq x}.
 \end{array}
\end{equation}
Note that we could have chosen any other of the $S_i$ (scaled by $2^\mathcal{L}$) as the root
simplex and we do not see any advantage or disadvantage in doing so.
\subsection{The type and Tet-id of a $d$-simplex}
\label{sec:type_id}
Making use of Property \ref{property:commref}, we define the following.
\begin{definition}
\label{def:type}
Each $d$-simplex $T\in\mathcal{T}_d$ of level $\ell$ lies in a $d$-cube of the hexahedral mesh
that is part of a uniform level $\ell$ refinement of $[0,2^{\mathcal{L}}]^d$.
This specific cube is the \emph{associated cube} of $T$ and denoted by $Q_T$.
The $d$-simplex $T$ is a scaled and shifted version of exactly one of the
six tetrahedra $S_i$ that constitute the unit cube, and
we define the \emph{type} of $T$ as this number, $\type(T):=i$.
\end{definition}
The anchor node of a subcube of level $\ell$ is the particular corner of
that cube with the smallest $x$-, $y$- (and $z$-) coordinates.
This means that for each simplex $T$ in the refinement from Figure \ref{fig:refineddiagram}
the anchor node of $T$ and the anchor node of its associated cube coincide.
Any two $d$-simplices in $\mathcal{T}_d$ with the same associated cube are distinguishable by their type.

From Bey's observation from Figure \ref{fig:refineddiagram} it follows that any simplex in $\mathcal T_d$
can be obtained by specifying a level $\ell$, then choosing one level $\ell$ subcube of the root cube and
finally fixing a type.
This provides motivation for the following definition.
\begin{definition}[Tet-id]
 For $T=[\vec x_0,\dots,\vec x_d]\in\mathcal{T}_d$ we define the \emph{Tet-id} of $T$ as the tuple of its anchor node
 and type; thus
 \begin{equation}
 \textrm{Tet-id}(T):=(\vec x_0,\type(T)).
 \end{equation}
 \end{definition}

 \begin{corollary}
 \label{cor:tetidunique}
 Let $T,T'\in\mathcal T_d$. Then $T=T'$ if and only if their Tet-ids and levels are the same.
 \end{corollary}

Note that in an arbitrary adaptive mesh there can be simplices with different levels and each 
simplex $T$ has an associated cube of level $\ell(T)$.
In particular, simplices with the same anchor node can have different associated cubes if their
levels are not equal.

 Since any simplex in $\mathcal T_d$ can be specified by the Tet-id and level, the Tet-id provides an im\-por\-tant tool
 for our work.
 The construction of the TM-index in the next section and the algorithms that
 we present in Section \ref{sec:algos} rely on the Tet-id as the fundamental
 data of a simplex.
 All information about a mesh can be extracted from the Tet-id and level of
 each element.

\begin{table}
\begin{center}
\raisebox{5.53ex}{
 \begin{tabular}{|rc|cccc|}
 \hline
 \multicolumn{2}{|c|}{{\mytabvspace Ct}}&\multicolumn{4}{c|}{Child}\\
 \multicolumn{2}{|c|}{2D}  &\mytabvspace $T_0$ & $T_1$ & $T_2$ & $T_3$\\[0.2ex] \hline
  \multirow{2}{*}{b}&0 & 0 & 0 & 0 & 1 \\
  &1 & 1 & 1 & 1 & 0 \\
  \hline
 \end{tabular}
 }
 \begin{tabular}{|rc|ccccc|}
 \hline
 \multicolumn{2}{|c|}{{\mytabvspace Ct}}&\multicolumn{5}{c|}{Child}\\
  \multicolumn{2}{|c|}{{\mytabvspace 3D}}
  & $T_0, \ldots, T_3$ & $T_4$ & $T_5$ & $T_6$ & $T_7$ \\[0.2ex] \hline
  \multirow{6}{*}{b}&0 & 0 & 4 & 5 & 2 & 1\\
  &1 & 1 & 3 & 2 & 5 & 0\\
  &2 & 2 & 0 & 1 & 4 & 3\\
  &3 & 3 & 5 & 4 & 1 & 2\\
  &4 & 4 & 2 & 3 & 0 & 5\\
  &5 & 5 & 1 & 0 & 3 & 4\\\hline
 \end{tabular}
 \end{center}
 \caption{For a $d$-simplex $T$ of type $b$ the table gives the types $\textrm{Ct}(T_0),\dots,\textrm{Ct}(T_{2^d-1})$ of $T's$ children.
 The corner-children $T_0,T_1,T_2$ (and in 3D also $T_3$) always have the same type as $T$.
 }
 \label{table:typesofchildren}
\end{table}

 Since the root simplex has type $0$, in a uniform refinement
 more simplices have type $0$ than any other type.
 However, a close examination of Table \ref{table:typesofchildren} together with a short inductive argument leads to the following proposition.
\begin{proposition}
 In the limit $\ell\rightarrow\infty$ the different types of simplices
 in a uniform level $\ell$ refinement of $\mathcal{T}_d$
 occur in equal ratios.
\end{proposition}

\subsection{Encoding of the tetrahedral Morton index}
\label{sec:TMindex}
In addition to the anchor coordinates the TM-index also depends on the types of
all ancestors of a simplex. In order to define the TM-index we
start by giving a formal
definition of the interleaving operation and some additional information.
\begin{definition}
 We define the \emph{interleaving} $a \ainter b$  of two $n$-tuples
 $a=$ $(a_{n-1},$ $\dots,a_0)$ and 
 $b=(b_{n-1},\dots,b_0)$  as the $2n$-tuple obtained
 by alternating the entries of $a$ and $b$:
 \begin{equation}
  a \ainter b \, {:=} \, (a_{n-1},b_{n-1},\dots,a_0,b_0).
 \end{equation}
 The interleaving of more than two $n$-tuples $a^1,\ldots,a^m$ is defined analogously as the $mn$-tuple
 \begin{equation}
a^1 \ainter \cdots \ainter a^m \, {:=} \, (a^1_{n-1},a^2_{n-1},\ldots,a^m_{n-1},a^1_{n-2},\ldots,a^{m-1}_0,a^m_0).
 \end{equation}
\end{definition}

\begin{remark}
The TM-index of a $d$-simplex $T\in\mathcal T_d$ that we are
going to define is constructed by interleaving $d+1$ $\mathcal L$-tuples, where
the first $d$ are the binary re\-pre\-sentations of the coordinates of $T$'s anchor
node and the last
is the tuple consisting of the types of the ancestors of $T$.
\end{remark}

\begin{definition}
\label{def:XYZ}
Let $T\in \mathcal{T}_3$ be a tetrahedron of refinement level $\ell$
 with anchor node $\vec x_0 = (x,y,z)^T\in\IL^3$.
 Since $x,y,z\abst{\in}\IN_{0}$ with $0\leq x,y,z < 2^\mathcal{L}$, we can express them as binary numbers with $\mathcal L$ digits, writing
 \begin{equation}
 x = \sum_{j=0}^{\mathcal{L}-1} x_j 2^j,\quad y = \sum_{j=0}^{\mathcal{L}-1} y_j 2^j,\quad z = \sum_{j=0}^{\mathcal{L}-1} z_j 2^j.
 \end{equation}
 We define the $\mathcal L$-tuples $X$, $Y$, and $Z$
 as the $\mathcal L$-tuples consisting of the binary digits of $x$, $y$, and $z$; thus,
 \begin{subequations}
 \begin{align}
 X = X(T)&:=(x_{\mathcal L -1},\dots,x_0),\\
 Y = Y(T)&:=(y_{\mathcal L -1},\dots,y_0),\\
 Z = Z(T)&:=(z_{\mathcal L -1},\dots,z_0).
 \end{align}
 \end{subequations}
 In 2D we get the same definitions with $X$ and $Y$, leaving out the $z$-coordinate.
\end{definition}

\begin{definition}\label{def:BvonT}
For a $T\in\mathcal T_d$ of level $\ell$ and each $0\leq j \leq \ell$ let $T^j$ be the (unique) ancestor of $T$ of level $j$. In particular, $T^\ell=T$.
We define $B(T)$ as the $\mathcal L$-tuple consisting of the types of $T$'s ancestors in the first $\ell$ entries, starting
with $T^1$. The last $\mathcal{L}-\ell$ entries of $B(T)$ are zero:
\begin{equation}
 B=B(T):=\left(\underbrace{\type(T^1),\type(T^2),\ldots,\type(T)}_{\ell\textrm{ entries}},0,\ldots,0\right),
\end{equation}
Thus, if we write $B$ as an $\mathcal{L}$-tuple with indexed entries $b_i$
\begin{equation}
B=B(T)=(b_{\mathcal L-1},\dots,b_0) \abst{\in} \set{0,\dots,d!-1}^\mathcal{L},
\end{equation}
then the $i$th entry $b_i$ is given as
\begin{equation}
 b_i= \left\lbrace\begin{matrix} \type(T^{\mathcal L -i}) & \mathcal L-1 \geq i \geq \mathcal L -\ell, \\[1ex]
                0      & \mathcal L -\ell > i \geq0.\hphantom{i-\ell}
                \end{matrix}\right.
\end{equation}
\end{definition}

\begin{definition}[tetrahedral Morton Index]\label{def:mortonindex}
We define the \emph{tetrahedral Morton index} (\emph{TM-index}) $m(T)$ of a $d$-simplex $T\in\mathcal{T}_d$ as the interleaving
of the $\mathcal L$-tuples $Z$ (for tetrahedra), $Y$, $X$ and $B$. Thus,
\begin{subequations}
 \label{eq:mortondef}
 \begin{equation}
   m(T)\abst{:=} Y\ainter X \ainter B
 \end{equation}
 for triangles and
 \begin{equation}
   m(T)\abst{:=} Z\ainter Y\ainter X \ainter B
 \end{equation}
 \end{subequations}
 for tetrahedra.
\end{definition}
This index resembles the well-known Morton index or Z-order for $d$-dimensional cubes, which we denote by $\widetilde m$ here.
For such a cube $Q$ the Morton index is usually defined as the bitwise interleaving of its coordinates.
Thus $\widetilde m(Q) = Z\inter Y\inter X$, respectively, $\widetilde m(Q) =
Y\inter X$; see \cite{Morton66,SundarSampathBiros08,BursteddeWilcoxGhattas11}.

As we show in Section \ref{sec:algos}, the TM-index can be computed from the Tet-id of $T$ with no further information given.
Thus, in an implementation it is not necessary to store the $\mathcal{L}$-tuple $B$.

The TM-index of a $d$-simplex builds up from packs of $d$ bits $z_i$ (for
tetrahedra), $y_i$, and $x_i$ followed by a type $b_i\in\set{0,\ldots,d!-1}$.
Since $d! = 2 < 4$ for $d=2$, we can interpret the 2D TM-index as
a quarternary number with digits $(y_ix_i)_2$ and $b_i$:
\begin{subequations}
\label{eq:mortonoctal}
\begin{equation}
\begin{array}{ccc}
\label{eq:mortonquater}
 m(T)  &=& ((y_{\mathcal L -1}x_{\mathcal L -1})_2,b_{\mathcal L -1},\ldots,(y_0x_0)_2,b_0)_4\\[1ex]
 &=& \displaystyle\sum_{i=0}^{\mathcal L-1}\left( (2y_i+x_i)4^{2i+1} + b_i4^{2i}\right).
\end{array}
\end{equation}
Similarly we can interpret it as an octal number with digits $(z_iy_ix_i)_2$ and $b_i$ for $d=3$, since then $d! = 6 < 8$:
\begin{equation}
\begin{array}{ccc}
 m(T) & = &((z_{\mathcal L -1}y_{\mathcal L -1}x_{\mathcal L -1})_2,b_{\mathcal L -1},\ldots,(z_0y_0x_0)_2,b_0)_8\\[1ex]
 &=& \displaystyle\sum_{i=0}^{\mathcal L-1}\left( (4z_i+2y_i+x_i)8^{2i+1} + b_i8^{2i}\right).
 \end{array}
\end{equation}
\end{subequations}
The entries in these numbers are only nonzero up to the level $\ell$ of $T$, since
$x_{\mathcal{L}-i} = y_{\mathcal{L}-i} = (z_{\mathcal{L}-i} = ) b_{\mathcal{L}-i} = 0$ for all $i>\ell$.
The octal/quarternary representation \eqref{eq:mortonoctal} directly gives an
order on the TM-indices, and therefore it is possible to construct an
SFC from it, which we examine further in Section
\ref{sec:SFC}.
We use $m(T)$ to denote both the $(d+1)\mathcal L$-tuple from \eqref{eq:mortondef} and the number given by \eqref{eq:mortonoctal}.

\begin{figure}
\center
\begin{minipage}{0.48\textwidth}
   \includegraphics[width=0.9\textwidth]{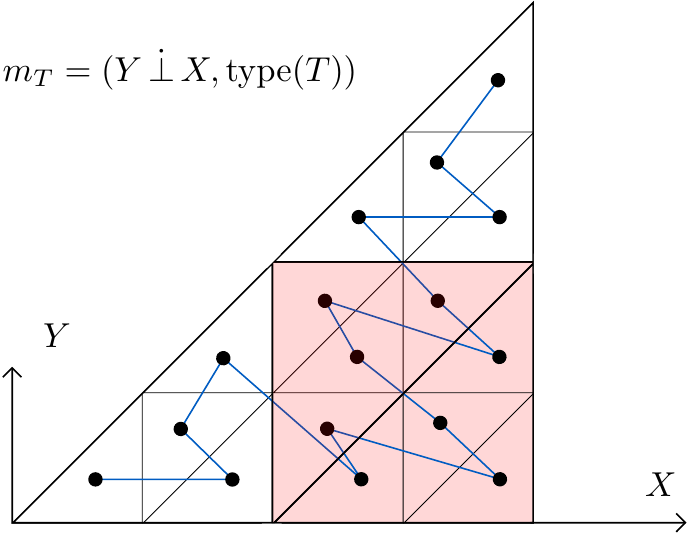}
\end{minipage}
\begin{minipage}{0.48\textwidth}
   \includegraphics[width=0.9\textwidth]{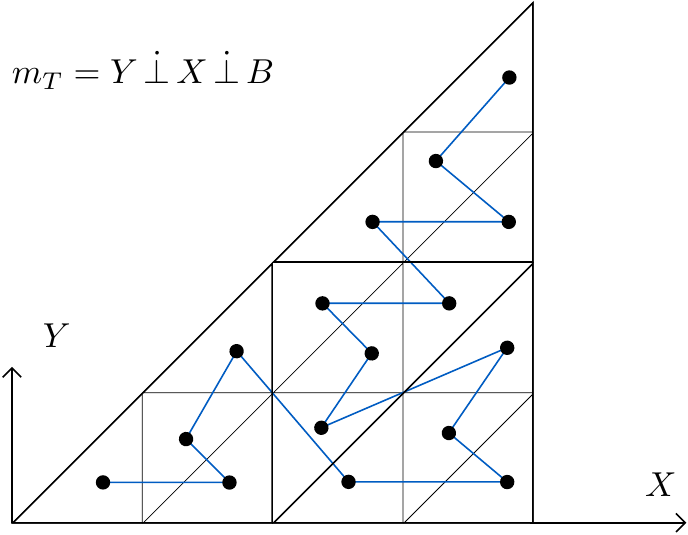}
\end{minipage}
   \caption{Comparing a straightforward definition of a Morton-type SFC
          with our approach.
          Left: The SFC arising from taking the Morton order of the quadrants
          and only dividing into triangles on the last level.
          Thus the index is  $(Y\inter X,\type(T))$. As we see
          on the two coarse triangles that are shaded, the children of a level 1 triangle are not necessarily
          traversed before any other triangle is traversed.
          Such a locality property would be desirable, and therefore this index is not suitable for our purposes.
          Right: The SFC arising from the TM-index from our Definition \ref{def:mortonindex}. We see
          that for each level 1 triangle its four children are traversed as a group.
          Theorem \ref{thm:IndexProps} states that this property holds for any parent triangle/tetrahedron.
          The order in which the children are traversed depends (only) on the type of the parent and is different from Bey's
          order given by \eqref{eq:childnumbers}.}
   \label{fig:mortonandwrongmorton}
\end{figure}

Let us look at Figure \ref{fig:mortonandwrongmorton} for a short example to motivate this definition of the TM-index.
Since the anchor coordinates and the type together with the level uniquely determine a $d$-simplex in $\mathcal T_d$,
one could ask why we do not define the index to be $((Z\ainter )Y\ainter X,\type(T))$,
a pair consisting of the Morton index of the associated cube of $T$ and the type of $T$.
This index was introduced for triangles in a slightly modified version as semiquadcodes in \cite{OtooZhu93}
and would certainly require less information since the computation of the sequence $B$ would not be necessary.
However, it  results in an SFC that traverses the leaf cubes of a refinement in the usual Z-order and inside of each cube it
traverses the $d!$ different simplices in the order $S_0,\ldots,S_{d!-1}$.
As a result, there can be simplices $T$ whose children are not traversed as a group,
which means that there is a tetrahedron $T'$,
which is not an ancestor or descendant of $T$, such that some child $T_i$ of $T$ is traversed before $T'$ and $T'$ is traversed before another child $T_j$ of $T$.
Theorem \ref{thm:IndexProps} states that this problem does not occur with the TM-index that we have defined above.
Figure \ref{fig:mortonandwrongmorton} compares the two approaches for a uniform level 2 refinement of $T^0_2$.

\subsection{A different approach to derive the TM-index}

\begin{figure}
   \center
   \def\svgwidth{0.35\textwidth}
   \includegraphics{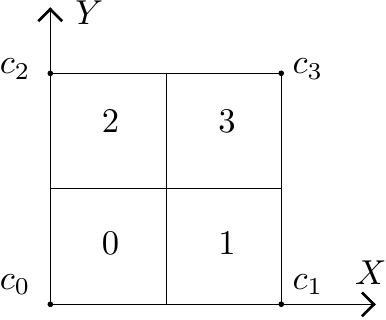}
   \hspace{10ex}
   \def\svgwidth{0.35\textwidth}
   \includegraphics{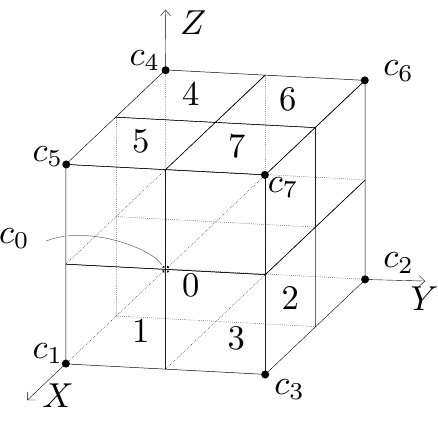}
    \caption{Left: A square is refined to four children, each of which corresponds to a corner of the square.
             The number of the corner is the cube-id of that child.
           Right: In three dimensions a cube is refined to eight children. Their cube-ids and corner numbers are shown as well.}
    \label{fig:cubeid}
\end{figure}

There is another interpretation of the TM-index, which is particularly useful for the AMR algorithms presented in Section \ref{sec:algos}.
In order to define it we introduce the concept of the so-called cube-id.
According to Figure \ref{fig:sechstetra} we number the $2^d$ corners of a  $d$-dimensional cube by $c_0,\ldots,c_{2^{d-1}}$ in a $zyx$-order ($x$ varies fastest).
When refining a cube to $2^d$ children, each child has exactly one of the $c_i$ as a corner, and
it is therefore convenient to number the children by $c_0,\ldots,c_{2^{d-1}}$ as well.
For the child $c_i$ we call the number $i$ the \emph{cube-id} of that child; see Figure \ref{fig:cubeid} for an illustration.
Since each cube $Q$ that is not the root has a unique parent, it also has a unique cube-id.
This cube-id can easily be computed by interleaving the last significant bits
of the $z$- (in 3D), $y$-, and $x$-coordinates of $Q$'s anchor node.
\begin{definition}
Because each $d$-simplex $T\abst{\in}\mathcal{T}_d$ of level $\ell$ has a unique associated cube we define
the \emph{cube-id} of $T$ to be the cube-id of the associated cube of $T$, that is, the $d$-cube of level
$\ell$ that has the same anchor node as $T$.
\end{definition}
If $X$, $Y$ (and $Z$) are as is in Definition \ref{def:XYZ} then we can write the cube-id of $T$'s ancestors as
\begin{equation}
\begin{array}{lc}
  \cid(T^i) = (y_ix_i)_2  & \textrm{in 2D},\\[1ex]
\cid(T^i) = (z_iy_ix_i)_2 & \textrm{in 3D},
\end{array}
\end{equation}
and therefore using \eqref{eq:mortonoctal} we can rewrite the TM-index of $T$ as
\begin{equation}\label{eq:mortonentries}
 m(T) = (\cid(T^1),\type(T^1),\dots,\cid(T^\ell),\type(T^\ell),0,\dots,0)_{2^d}.
\end{equation}
This resembles the Morton index of the associated cube $Q_T$ of $T$, since we can write this as
\begin{equation}
 \widetilde m({Q_T}) = (\cid(Q^1),\dots,\cid(Q^\ell),0,\dots,0)_{2^d}.
\end{equation}

\subsection{Properties of the TM-index}\label{sec:Prop}
In this section we show that the TM-index has properties similar to that of the
Morton index for cubes.  We are particularly in\-te\-rested in locality
properties stating that refining a simplex only locally changes the order
given by the TM-index.
We begin by stating the uniqueness of the TM-index if additionally a level $\ell$ is given.
\begin{proposition}\label{prop:uniquetet}
Together with a refinement level $\ell$, the TM-index $m(T)$ 
u\-nique\-ly determines a $d$-simplex in $\mathcal T_d$.
\begin{proof}
If $\ell=0$, then there is only one simplex of level $\ell$ in $\mathcal T_d$, which is $T^0_d$.
So let $\ell>0$ and $m=m(T)$ be given as in \eqref{eq:mortondef}, and let $\ell$ be the level of $T$.
From $m$ we can compute the $x$- ,$y$- (and $z$-)
coordinates of the associated cube of $T$.
We can also compute the type of $T$ from the TM-index.
By Corollary \ref{cor:tetidunique} this information uniquely determines $T$.
\end{proof}
\end{proposition}
For the Morton index $\widetilde m$ for cubes
the following important properties are known\cite{SundarSampathBiros08}:
\begin{enumerate}[(i)]
 \item The Morton indices of the descendants of a parent cube are larger than or equal to the index of the parent cube.
 \item A Morton index of a cube $Q$ is the prefix of an index of a cube $P$ of higher level than $Q$ if and only if
       $P$ is a descendant of $Q$.
 \item Refining only changes the SFC locally. Thus, if $Q$ is a cube and $P$ is a cube with $\widetilde m(Q)<\widetilde m(P)$
       and $P$ is not a descendant of $Q$, then $\widetilde m({Q'})<\widetilde m(P)$ for each descendant $Q'$ of $Q$.
\end{enumerate}
Property (iii) defines a hierarchic invariant of the SFC
that is specific to our construction (see Figure~\ref{fig:mortonandwrongmorton}).
We show below that properties (i), (ii) and (iii) hold for $d$-simplices and
the TM-index described by \eqref{eq:mortondef}.
\begin{theorem}\label{thm:IndexProps}
 For arbitrary $d$-simplices $T\neq S\in\mathcal T_d$ the TM-index satisfies
the following:
\begin{enumerate}[(i)]
 \item  If $S$ is a descendant of $T$ then $m(T)\leq m({S})$.
 \item  If $\ell(T) < \ell(S)$, then $m(T)$ is a prefix of $m(S)$ if and only if $S$ is a descendant of $T$.
 \item  If $m(T)<m(S)$ and $S$ is no descendant of $T$, then for each descendant $T'$ of $T$
we have
\begin{equation}
 m(T)\leq m({T'}) < m(S).
\end{equation}
\end{enumerate}
\end{theorem}
The proof of Theorem \ref{thm:IndexProps} requires some work and we need to show a technical result first.
Hereby, we consider only the 3D case, since for 2D the argument is completely analogous.
We define an embedding of the set of all TM-indices into the set of  Morton
indices for 6D cubes. Since the properties (i)--(iii) hold for
these cubes it follows that they hold
for tetrahedra as well.
To this end, for a given tetrahedron $T\abst{\in}\mathcal{T}_3$ we interpret each entry $b_j$ of $B(T)$ as a $3$-digit binary number
\begin{equation}
 b_j \abst{=} (b_j^2\, b_j^1\, b_j^0)_2,
\end{equation}
which is possible since $b_j\,{\in}\, \set{0,\dots,5}$.
We obtain three new $\mathcal{L}$-tuples $B^2, B^1, B^0$ satisfying
\begin{equation}
 B\abst{=} B^2\ainter B^1\ainter B^0 ,
\end{equation}
and thus we can rewrite the TM-index as
\begin{equation}
\label{eq:rewritemorton}
 m(T)\abst{=} Z\ainter Y\ainter X \ainter B^2\ainter B^1\ainter B^0.
\end{equation}
Note that we can interpret each $B^i$ as an $\mathcal L$-digit binary number
for which we have $0\leq B^i<2^\mathcal{L}$.
Now let $\mathcal{Q}$ denote the set of all 6D cubes that are a child of the cube $Q_0 \, {:=} \, [ 0, 2^\mathcal{L}]^6$:
\begin{equation}
 \mathcal{Q}\abst{=} \set{Q \abst{|} Q \textnormal{ is a descendant of } Q_0
 \textnormal{ of level } 0\leq \ell\leq\mathcal{L}}.
\end{equation}
Since a cube $Q\,{\in}\, \mathcal{Q}$ is uniquely determined by the six coordinates $(x_0,\dots,x_5)$ of its anchor node plus its level $\ell$,
we also write
$Q\abst{=} Q_{(x_0,\dots,x_5),\ell}$.
Note that the Morton index for a cube can be defined as the bitwise interleaving of its anchor node coordinates \cite{Morton66}:
\begin{equation}
 \widetilde{m}(Q)\abst{=} X^5\ainter X^4\ainter X^3\ainter X^2\ainter X^1\ainter X^0.
\end{equation}

\begin{proposition}
\label{prop:mapfromTtoQ}
The map
 \begin{equation}\label{eq:MapTQ}
  \begin{array}{rccl}
   \Phi\colon&\mathcal{T}_3 & \longrightarrow & \mathcal{Q}, \\
   &T & \longmapsto & Q_{(B^0(T),B^1(T),B^2(T),x(T),y(T),z(T)),\ell(T)} \\
  \end{array}
 \end{equation}
 is injective and satisfies
 \begin{equation}
   \widetilde m({\Phi(T)})\abst{=} m(T).
 \end{equation}
 Furthermore, it fulfills the property that $T'$ is a child of $T$ if and only if $\Phi(T')$ is a child of $\Phi(T)$.
 \begin{proof}
  The equation $m(T)=\widetilde m({\Phi(T))}$ follows directly from the definitions of the TM-indices on $\mathcal T_3$ and $\mathcal Q$.
  From Lemma \ref{prop:uniquetet} we conclude that $\Phi$ is injective.
  Now let $T',T\abst{\in}\mathcal{T}_3$, where $T'$ is a child of $T$.
  Furthermore, let $\ell=\ell(T)$.
  We know that $Q':=\Phi(T')$ is a child of $Q:=\Phi(T)$ if and only if for each $i\in\set{0,\dots,5}$
  it holds that
  \begin{equation}\label{eq:xiequation}
   x_i(Q')\in\set{x_i(Q),x_i(Q)+2^{\mathcal L -(\ell+1)}}.
  \end{equation}
  Because of the underlying cube structure (compare Figure~\ref{fig:refineddiagram}) we know that the $x$-coordinate
  of the anchor node of $T'$ satisfies
  \begin{equation}
   x(T')\in\set{x(T),x(T)+2^{\mathcal L -(\ell+1)}},
  \end{equation}
  and likewise for $Y(T')$ and $Z(T')$.
  Therefore, \eqref{eq:xiequation} holds for $i=3,4,5$.
  By definition $B^j(T')$ is the same as $B^j(T)$ except at position $\mathcal L-(\ell+1)$, where
   \begin{equation}
    B^j(T')_{\mathcal L-(\ell+1)} = b_{\mathcal L-(\ell+1)}^j(T')\in\set{0,1}
   \end{equation}
   and
   \begin{equation}
  B^j(T)_{\mathcal L-(\ell+1)} = 0.
   \end{equation}
   Hence, we conclude that \eqref{eq:xiequation} also holds for $i=0,1,2$.
   So $\Phi(T')$ is a child of $\Phi(T)$.

   To show the other implication, let us assume that
   $\Phi(T')$ is a child of $\Phi(T)$.
   Since $\ell(T')=\ell(\Phi(T'))>0$,  $T'$ has a parent and
   we denote it by $P$.
   In the argument above we have shown that $\Phi(P)$ is the parent of $\Phi(T')$
   and because each cube has a unique parent
   the identity $\Phi(P)=\Phi(T)$ must hold. Therefore, we get $P=T$ since $\Phi$ is injective; thus, $T'$ is the child of $T$.
\end{proof}
\end{proposition}
Inductively we conclude that $T'$ is a descendant of $T$ if and only if $\Phi(T')$ is a de\-scen\-dant of $\Phi(T)$.
Now Theorem \ref{thm:IndexProps} follows, because the desired properties (i)--(iii) hold for the Morton index of cubes.
\qed

\subsection{The space-filling curve associated to the TM-index}
\label{sec:SFC}
By interpreting the TM-indices as $2^d$-ary numbers as in \eqref{eq:mortonoctal}
we get a total order on the set of all possible TM-indices,
and therefore it gives rise to an SFC for any refinement $\mathscr{S}$ of $T^0_d$.
In this section we further examine the SFC derived from the TM-index.
We give here a recursive description of it,
similarly to how it is done for the Sierpinski curve and other
cubical SFC by Haverkort and van Walderveen \cite{HaverkortWalderveen10}.

Part (iii) of Theorem \ref{thm:IndexProps} tells us that the descendants of a simplex $T$ are traversed
before any other simplices with a higher TM-index than $T$ are traversed.
However, the order that the children of $T$ have relative to each other can be different to the
order of children of another simplex $T'$.
In particular the order of the simplices defined by the TM-index differs from the order
\eqref{eq:childnumbers} defined by Bey.
We observe this behavior in 2D in Figure~\ref{fig:mortonandwrongmorton} on the right-hand side:
For the level 1 triangles of type $0$ the children are traversed in the order
\begin{equation}
 T_0, T_1, T_3, T_2
\end{equation}
and the children of the level 1 triangle of type $1$ are traversed in the order
\begin{equation}
 T_0, T_3, T_1, T_2.
\end{equation}
In fact, the order of the children of a simplex $T$ depends only on the type of $T$, as
we show in the following Proposition.
\begin{proposition}
\label{proposition:locindexdeptype}
 If $T,T'\in \mathcal T_d$ are two $d$-simplices of given type $b = \type(T)=\type(T')$,
 then there exists a unique permutation $\sigma \equiv \sigma_b$ of
 $\set{0,\dots,2^d-1}$ such that
 \begin{equation}
 \begin{array}{c}
  m(T_{\sigma(0)})<m(T_{\sigma(1)})<\dots<m(T_{\sigma(2^d-1)}), \\[1ex]
  \textrm{and}\\[1ex]
  m(T'_{\sigma(0)})<m(T'_{\sigma(1)})<\dots<m(T'_{\sigma(2^d-1)}).
 \end{array}
 \end{equation}
 Thus, the children of $T$ and the children of $T'$ are in the same order with respect to their TM-index.
 \begin{proof}
 By ordering the children of $T$ and $T'$ with respect to their TM-indices,
 we obtain $\sigma$ and $\sigma'$ with
 \begin{equation}
 \begin{array}{c}
  m(T_{\sigma(0)})<m(T_{\sigma(1)})<\dots<m(T_{\sigma(2^d-1)}), \\[1ex]
  m(T'_{\sigma'(0)})<m(T'_{\sigma'(1)})<\dots<m(T'_{\sigma'(2^d-1)}).
 \end{array}
 \end{equation}
 These permutations are well-defined and unique with this property because different
 simplices of the same level never have the same TM-index; see Proposition
 \ref{prop:uniquetet}.
 It remains to show that $\sigma'=\sigma$.
Let $\ell=\ell(T)$ and $\ell'=\ell(T')$.
Since the TM-indices of the children of $T$ do all agree up to level $\ell$, we see,
using the notation from \eqref{eq:mortonoctal}, that their order $\sigma$ depends
only on the $d+1$ numbers ($z$ is omitted for $d=2$)
\begin{equation}
z_{\mathcal L-(\ell+1)}(T_i),\quad y_{\mathcal L-(\ell+1)}(T_i),\quad x_{\mathcal L-(\ell+1)}(T_i)\quad\textrm{and}\quad b_{\mathcal L-(\ell+1)}(T_i).
\end{equation}
The same argument applies to $\sigma'$ and $\ell'$.
From now on we carry out the computations for $d=3$.
Since $\type(T)=\type(T')$ we can write
\begin{equation}
 T = \lambda T' + \vec c,
\end{equation}
with
\begin{equation}
\lambda = 2^{\ell'-\ell},\quad\vec c = \begin{pmatrix}
                    x(T)-x(T')\\[0.6ex]
                    y(T)-y(T')\\[0.6ex]
                    z(T)-z(T')
                                  \end{pmatrix}.
\end{equation}
Since the refinement rules \eqref{eq:childnumbers} commute with scaling and translation we also obtain
\begin{equation}
 T_i = \lambda T'_i + \vec c
\end{equation}
for the children of $T$ and $T'$
and therefore
\begin{equation}
 b_{\mathcal L-(\ell+1)}(T_i)=\type(T_i)=\type(T'_i)=b_{\mathcal L-(\ell'+1)}(T'_i)
\end{equation}
for $0\leq i < 2^d$.
Furthermore, 
we have
\begin{equation}
x_{\mathcal L-(\ell+1)}(T_i) = (x(T_i)-x(T))2^{-\mathcal{L}+(\ell+1)}
\end{equation}
from which we derive
\begin{equation}
 \begin{array}{ccl}
  x_{\mathcal L-(\ell+1)}(T_i)&=& \lambda(x(T_i') - x(T')) 2^{-\mathcal{L}+(\ell+1)}\\[0.8ex]
  &=& 2^{\ell'-\ell} (x(T_i') - x(T'))
  2^{-\mathcal{L}+(\ell+1)} \\[0.8ex]
  &=& (x(T_i') - x(T')) 2^{-\mathcal{L}+(\ell'+1)} \\[0.8ex]
  &=& x_{\mathcal L-(\ell'+1)}(T'_i),
 \end{array}
\end{equation}
and analogously
\begin{equation}
\begin{array}{ccc}
 y_{\mathcal L-(\ell+1)}(T_i) &=&y_{\mathcal L-(\ell'+1)}(T'_i), \\[0.8ex]
 z_{\mathcal L-(\ell+1)}(T_i) &=&z_{\mathcal L-(\ell'+1)}(T'_i).
\end{array}
\end{equation}
This shows that the tetrahedral Morton order of the children of $T$ and $T'$ are the same and $\sigma'$ must equal $\sigma$.
\end{proof}
\end{proposition}

\begin{figure}
   \center
   \def\svgwidth{0.9\textwidth}
   \includegraphics{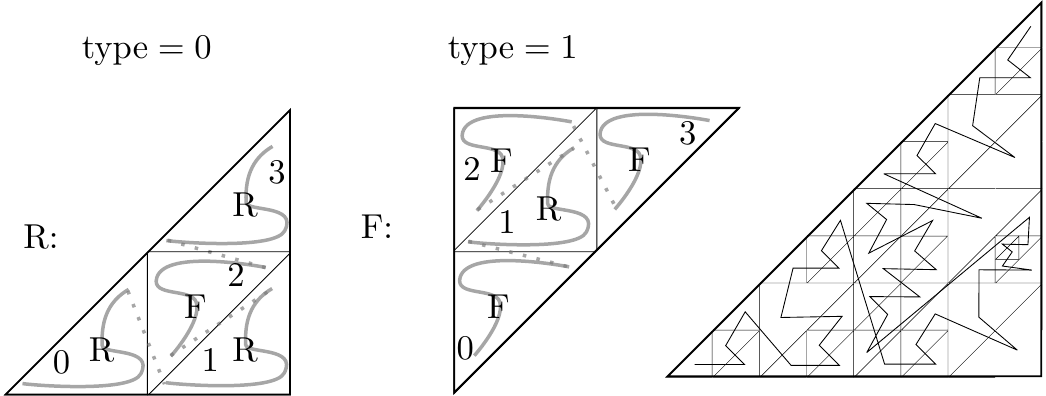}
    \caption{Left: Using the notation from \cite{HaverkortWalderveen10} we
recursively describe the SFC arising from the TM-index for triangles. The
number inside each child triangle
        is its local index.
        $R$ denotes the refinement scheme for type 0 triangles and $F$ for type 1 triangles.
        This pattern can be obtained from Tables \ref{table:typesofchildren} and \ref{table:BeytoIndex}.
        Right: The SFC for an example adaptive refinement of
        the root triangle.}
    \label{fig:haverkortSFC}
\end{figure}

\begin{definition}
 \label{def:localindex}
 Let $T\in\mathcal T_d$ such that $T$'s parent $P$ has type $b$ and $T$ is the $i$th child of $P$
 according to Bey's order \eqref{eq:childnumbers},
  $0\leq i<2^d$. We call the number $\sigma_b(i)$ the \emph{local index} of
 the $d$-simplex $T$ and use the notation
 \begin{equation}
   \label{eq:Ilocsigma}
I_\mathrm{loc}(T):=\sigma_b(i)
 \end{equation}
 to denote the child number in the TM-ordering, subsequently written TM-child.
 By definition, the local index of the root simplex is zero, $I_\mathrm{loc}(T^0_d):=0$.
 Table \ref{table:BeytoIndex} lists the local indices for each parent type.
\end{definition}
Thus, we know for each type $0\leq b<d!$ how the children of a tetrahedron of type $b$ are traversed.
This gives us an approach for describing the SFC arising from the TM-index in a recursive fashion \cite{HaverkortWalderveen10}.
By specifying for each possible type $b$ the order and types of the children of a type $b$ simplex, we can build up the SFC.
In Figure \ref{fig:haverkortSFC} 
we describe the SFC for triangles in this way.
In three dimensions it is not convenient to draw the six pictures for the different types, but
the SFC can be derived similarly from Tables \ref{table:typesofchildren} and \ref{table:BeytoIndex}.
\begin{table}
\centering
\raisebox{6ex}{
\begin{tabular}{|rc|cccl|}
\hline
\multicolumn{2}{|c|}{\mytabvspace$I_\mathrm{loc}$}&\multicolumn{4}{c|}{Child}\\
 \multicolumn{2}{|c|}{2D}  &\mytabvspace  $T_0$ & $T_1$ & $T_2$ & $T_3$ \\[0.2ex]\hline
 \multirow{2}{*}{b}&\mytabvspace0 & 0 & 1 & 3 & 2    \\[0.2ex]
 &1 & 0 & 2 & 3 & 1   \\ \hline
\end{tabular}
}
\begin{tabular}{|rc|cccccccl|}
\hline
\multicolumn{2}{|c|}{\mytabvspace$I_\mathrm{loc}$}&\multicolumn{8}{c|}{Child}\\
\multicolumn{2}{|c|}{3D}&
\mytabvspace   $T_0$ & $T_1$ & $T_2$ & $T_3$ & $T_4$ & $T_5$ & $T_6$ & $T_7$\\[0.2ex]\hline
 \multirow{6}{*}{b}&\mytabvspace0 & 0 & 1 & 4 & 7 & 2 & 3 & 6 & 5   \\[0.2ex]
 &1 & 0 & 1 & 5 & 7 & 2 & 3 & 6 & 4   \\[0.2ex]
 &2 & 0 & 3 & 4 & 7 & 1 & 2 & 6 & 5   \\[0.2ex]
 &3 & 0 & 1 & 6 & 7 & 2 & 3 & 4 & 5   \\[0.2ex]
 &4 & 0 & 3 & 5 & 7 & 1 & 2 & 4 & 6   \\[0.2ex]
 &5 & 0 & 3 & 6 & 7 & 2 & 1 & 4 & 5   \\ \hline
\end{tabular}
\caption{The local index for the children of a $d$-simplex $T$ according to the
  TM-ordering.
  For each type $b$, the $2^d$ children $T_0,\dots,T_{2^d-1}$ of
      a simplex of this type can be ordered according to their TM-indices.
      The position of $T_i$ according to the TM-order is the local index $I_\mathrm{loc}(T_i) = \sigma_b (i)$.}
\label{table:BeytoIndex}
\end{table}

\section[Algorithms]{Algorithms on simplices}
\label{sec:algos}
In this section we present fundamental algorithms that operate on $d$-simplices in $\mathcal{T}_d$.
These algorithms include computations of parent and child simplices, computation of face-neighbors and computations
involved with the TM-index.
To simplify the notation we carry out all algorithms for tetrahedra and then describe how to modify them for triangles.
We introduce the data type \Tet{}
and do not distinguish between the abstract concept
of a \Tet{} and the geometric object (tetrahedron or triangle) that it represents.
The data type \Tet{} $T$ has the following members:
\begin{itemize}
 \item $T.\ell$ --- the refinement level of $T$;
 \item $T.\vec{x}=(T.x,T.y,T.z)$ --- the $x$- ,$y$- and $z$-coordinates of $T$'s anchor node,
    also sometimes referred to as $T.x_0$, $T.x_1$, and $T.x_2$;
 \item $T.b$ --- the type of $T$.
 \end{itemize}
 In 2D computations the parameter $T.z$ is not present.
 To avoid confusion we use the notation $\vec{x}$ to denote vectors in $\IZ^d$ and $x$ (without arrow) for
 integers, thus numbers in $\IZ$.
From Corollary \ref{cor:tetidunique} we know that the values stored in a \Tet{}
suffice to uniquely identify a $d$-simplex $T\abst{\in}\mathcal{T}$.
\begin{remark}[Storage requirement]
The algorithms that we present in this section only need this data as input for
a simplex resulting in a fixed storage size per \Tet.
If, for example, the maximum level $\mathcal{L}$ is 32 or less, then the
coordinates can be stored in one $4$-byte integer per dimension, while the
level and type occupy one byte each, leading to a total storage of
\begin{equation}
\begin{array}{cccr}
2 \times 4+1+1 &=& 10 &\textrm{bytes per \Tet{} in 2D,}\\
3 \times 4+1+1 &=& 14 &\textrm{bytes per \Tet{} in 3D.}
\end{array}
\end{equation}
\end{remark}
\begin{remark}[Runtime]
Most of these algorithms run in constant time independent of the maximum level $\mathcal L$.
The only operations
using a loop over the level
$\mathcal L$ or $T.\ell$, thus having $\mathcal{O}(\mathcal{L})$ runtime, are
computing the consecutive index from a \Tet{} and initializing a \Tet{}
according to a given consecutive index.
Hence, we show how to replace repetitive calls of these relatively
involved algorithms by more efficient constant-time ones.
\end{remark}

\subsection{The coordinates of a $d$-simplex}
The coordinates of the $d+1$ nodes of a $d$-simplex $T$
can be obtained easily from its Tet-id, the relation \eqref{eq:coordsofSb}, and simple
arithmetic:
 If $T$ is a $d$-simplex of level $\ell$, type $b$ and anchor node $\vec{x}_0\in\IZ^d$, then
 \begin{equation}
 \label{eq:TcoordfromS}
  T = 2^{\mathcal{L}-\ell}S_b + \vec{x}_0.
 \end{equation}
Hence, in order to compute the coordinates of $T$ we can take the coordinates of $S_b$, as given in \eqref{eq:coordsofSb},
and then use relation \eqref{eq:TcoordfromS}.
A closer look at \eqref{eq:coordsofSb} reveals that it is not necessary to examine all coordinates of $S_b$
in order to compute the $x_i$, but that they can also be computed arithmetically.
This computation is carried out in Algorithm \ref{alg:coordsofT}.

\begin{algorithm}
\caption{\texttt{Coordinates}(\aTet{} $T$)}
\label{alg:coordsofT}
$X\gets (T.\vec{x},0,0,0)$\;
$h\gets 2^{\mathcal{L}-\ell}$\;
$i\gets \lfloor\frac{T.b}{2}\rfloor$\Comment{Replace with $i\gets T.b$ for 2D}
\eIf{$T.b\algomod 2 = 0$}
{$j\gets (i+2) \algomod 3$}
{$j\gets (i+1) \algomod 3$}
$X[1]\gets X[0]+he_i$\;
$X[2]\gets X[1]+he_j$\Comment{Replace with $X[2]\gets X[0] + (h,h)^T$ for 2D}
$X[3]\gets X[0]+(h,h,h)^T$\Comment{Remove this line for 2D}
\Return $X$\;
\end{algorithm}

\subsection{Parent and child}
\label{sec:Parent}
\begin{figure}
\flushleft
\raisebox{5.6ex}{
\begin{minipage}{0.21\textwidth}
\center
\scalebox{0.9}{
\begin{tabular}{|rc|cc|}
 \hline
 \multicolumn{2}{|c|}{\mytabvspace $\mathrm{Pt}(c,b)$} &\multicolumn{2}{c|}{b}\\
 \multicolumn{2}{|c|}{2D}
 &\mytabvspace  0 & 1 \\[0.5ex]\hline
 \multirow{4}{*}{$c$}&\mytabvspace0& 0 & 1   \\[0.2ex]
 &1& 0 & 0   \\[0.2ex]
 &2& 1 & 1   \\[0.2ex]
 &3& 0 & 1   \\[0.2ex]\hline
\end{tabular} 
}
\end{minipage}
}
\begin{minipage}{0.41\textwidth}
\center
\scalebox{0.9}{
\begin{tabular}{|rc|cccccc|}
 \hline
 \multicolumn{2}{|c|}{\mytabvspace $\mathrm{Pt}(c,b)$} &\multicolumn{6}{c|}{b}\\
 \multicolumn{2}{|c|}{3D}&
 \mytabvspace
   0 & 1 & 2 & 3 & 4 & 5 \\[0.5ex]\hline
 \multirow{8}{*}{$c$}&\mytabvspace0& 0 & 1 & 2 & 3 & 4 & 5  \\[0.2ex]
 &1& 0 & 1 & 1 & 1 & 0 & 0  \\[0.2ex]
 &2& 2 & 2 & 2 & 3 & 3 & 3   \\[0.2ex]
 &3& 1 & 1 & 2 &  \cfbox{Magenta}{\color{blue}2}& 2 & 1   \\[0.2ex]
 &4& 5 & 5 & 4 & 4 & 4 & 5   \\[0.2ex]
 &5& 0 & 0 & 0 & 5 & \cfbox{Dandelion}{\color{PineGreen}5} & 5   \\[0.2ex]
 &6& 4 & 3 & 3 & 3 & 4 & 4  \\[0.2ex]
 &7& 0 & 1 & 2 & 3 & 4 & 5 \\ \hline
\end{tabular} 
}

\end{minipage}
\begin{minipage}{0.34\textwidth}
   \includegraphics[width=0.9\textwidth]{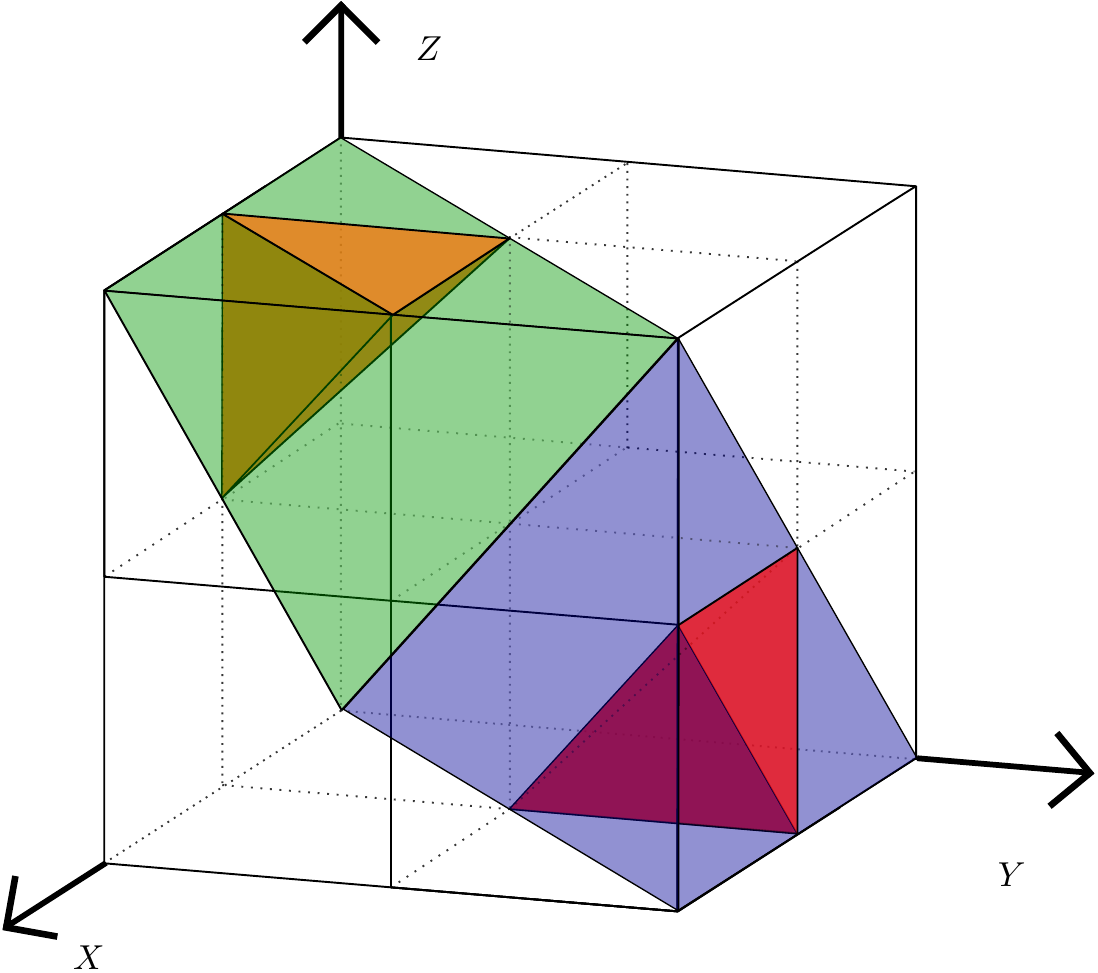}
\end{minipage}
   \caption{The type of the parent of a $d$-simplex $T$ can be determined from $T$'s cube-id $c$ and type $b$.
             Left: The values of $Pt$ from \eqref{eq:Pt} in the 2D case.
             Middle: The values of the function $Pt$ in the 3D case.
             Right: Two examples showing the computation in 3D. (1) The small tetrahedron in the top left corner (orange) has cube-id $5$ and type $4$,
              and its parent (green) can be seen to have type $5$.
              (2) The small tetrahedron at the bottom right (red) has cube-id $3$ and type $3$, and its parent (blue)
              has type $2$.
}
   \label{fig:parenttype}
\end{figure}

In this section we describe how to compute the Tet-ids of the parent $P(T)$ and
of the $2^d$ children $T_i$, $0\leq i<2^d$, of a given $d$-simplex $T\in\mathcal
T_d$.  Computing the anchor node coordinates of the parent is easy, since their
first $T.\ell-1$ bits correspond to
the coordinates of $T$'s anchor node and the rest of their bits is zero.
For computing the type of $P(T)$, we need the function
\begin{equation}
\label{eq:Pt}
 \begin{array}{rcl}
\textrm{Pt}\colon \set{0,\ldots,2^d-1}\times\set{0,\ldots,d!-1} &\longrightarrow &\set{0,\ldots,d!-1} ,\\[.5ex]
(\textrm{cube-id}(T),T.b)&\longmapsto & P.b,
\end{array}
\end{equation}
giving the type of $T$'s parent in dependence of its cube-id and type.
In Figure \ref{fig:parenttype} we list all values of this function for $d\in\set{2, 3}$.

\begin{algorithm}
 \caption{\texttt{c-id}(\aTet{} $T$,\Int $\ell$)%
}
 \label{alg:cube-id}
 \DontPrintSemicolon
 $i\gets0$, $h\gets 2^{\mathcal{L}-\ell}$\;
 $i\aorgl (T.x\bitwand h) \abst{?} 1 \colon 0$\;
 $i\aorgl (T.y\bitwand h) \abst{?} 2 \colon 0$\;
 $i\aorgl (T.z\bitwand h) \abst{?} 4 \colon 0$\Comment{Remove this line for 2D}
 \Return $i$
\end{algorithm}

\begin{algorithm}
 \caption{\texttt{Parent}(\aTet{} $T$)}
 \label{alg:parent}
 \DontPrintSemicolon
 $h\gets 2^{\mathcal{L}-T.\ell}$\;
 $P.\ell\gets T.\ell-1$\;
 $P.x\gets T.x \bitwand \neg h$\;
 $P.y\gets T.y \bitwand \neg h$\;
 $P.z\gets T.z \bitwand \neg h$\Comment{Remove this line for 2D}
 $P.b\gets \mathrm{Pt}(\texttt{c-id}(T,T.\ell),T.b)$\Comment{See \eqref{eq:Pt} and Figure \ref{fig:parenttype} for Pt}
 \Return $P$\;
\end{algorithm}
The algorithm \texttt{Parent} to compute the parent of $T$ now puts these two ideas together, computing the coor\-di\-nates and type of $P(T)$.
Algorithm \ref{alg:parent} shows an implementation. It uses Algorithm \ref{alg:cube-id} to compute the cube-id of a $d$-simplex.

For the computation of one child $T_i$ of $T$ for a given $i\in\set{0,\dots,2^d-1}$
we look at Bey's definition of the subsimplices in \eqref{eq:childnumbers} and see that in order to compute the anchor node
of the child we need to know some of the node coordinates $\vec x_0,\dots,\vec x_d$ of the parent simplex $T$.
These can be obtained via Algorithm \ref{alg:coordsofT}.
However, it is more efficient to compute only those coordinates
of $T$ that are actually necessary.
To compute the Tet-id of $T_i$ we also need to know its type.
The type of $T_i$ depends only on the type of $T$, and in the algorithm
we use the function \texttt{Ct} (children type) to compute this type.
\texttt{Ct} is effectively an evaluation of Table \ref{table:typesofchildren}.
Algorithm \ref{alg:children} shows now how to compute the coordinates of the $i$th child of $T$ in Bey's order.

When we would like to compute the $i$th child of a $d$-simplex $T$ of type $b$ with respect to the tetrahedral Morton order
(thus the child $T_k$ of $T$ with $I_\mathrm{loc}(T_k)=i$) we just call Algorithm \ref{alg:children}
with $\sigma_b^{-1}(i)$ as input.
The permutations $\sigma_b^{-1}$ are available from Table \ref{table:BeytoIndex};
see \eqref{eq:Ilocsigma} and Algorithm \ref{alg:TMchild}.

\begin{algorithm}
 \caption{\texttt{Child}(\aTet{} $T$,\Int $i$)}
 \label{alg:children}
 \DontPrintSemicolon
 $X\gets$ \texttt{Coordinates}($T$)\;
 \lIf{$i=0$}{$j\gets0$}

 \lElseIf{$i\in\set{1,4,5}$}{$j\gets1$}

 \lElseIf{$i\in\set{2,6,7}$}{$j\gets2$}

 \lElseIf(\IfComment{If $i={3}$\textbf{ then }$j\gets 1$ for 2D}){$i={3}$}{$j\gets3$}
 $T_i.\vec{x}\gets\frac{1}{2}(X[0]+X[j])$\;
 $T_i.b\gets$ \texttt{Ct}$(T.b,i)$\Comment{See Table \ref{table:typesofchildren}}
\end{algorithm}

\begin{algorithm}
  \caption{\texttt{TM-Child}(\aTet{} $T$,\Int $i$)}
  \label{alg:TMchild}
  \DontPrintSemicolon
  \Return \texttt{Child} ($T$, $\sigma_{T.b}^{-1} (i)$)
  \Comment{See Table~\ref{table:BeytoIndex}}
\end{algorithm}

\subsection{Neighbor simplices}

Many applications---e.g., finite element me\-thods---need to gather information about the face-neighboring simplices of
a given simplex in a refinement.
In this section we describe a level-independent constant-runtime algorithm to compute the
Tet-id of the same level neighbor along a specific face $f$ of a given $d$-simplex $T$.
This algorithm is very lightweight since it only requires a few arithmetic
computations involving the Tet-id of $T$ and the number $f$.
In comparison to other approaches to finding neighbors in constant time
\cite{LeeDeSamet01,BrixMassjungVoss11},
our algorithm does not involve the computation of any of $T$'s ancestors.

The $d+1$ faces of a $d$-simplex $T=[\vec x_0,\dots,\vec x_d]$ are numbered  $f_0,\dots,f_d$ in such a way that face $f_i$ is the face
not containing the node $\vec x_i$.
To examine the situation where two $d$-simplices of the same level share a common face,
let $\mathcal{T}^\ell_d$ denote a uniform refinement of $T^0_d$ of a given level $0\leq \ell \leq \mathcal{L}$,
\begin{equation}
 \mathcal{T}_d^\ell\, {:=} \, \set{T \abst{|} T \textnormal{ is a descendant of } T^0_d
 \textnormal{ of level } \ell}\abst{\subset} \mathcal{T}_d.
\end{equation}
$\mathcal{T}_d$ can be seen as the disjoint union of all the $\mathcal T^\ell_d$:
\begin{equation}
 \mathcal T_d = \bigcup_{\ell=0}^{\mathcal L} \mathcal T^\ell_d.
\end{equation}
Given a $d$-simplex $T\abst{\in} \mathcal{T}^\ell_d$ and a face number
$i\abst{\in}\set{0,\dots,d}$, denote $T$'s neighbor in $\mathcal{T}^\ell_d$
across face $f=f_i$ by $\mathcal{N}_{f}(T)$, and denote the face number of the
neighbor simplex $\mathcal{N}_f(T)$ across which $T$ is its neighbor by
$\tilde{f}(T)$. Hence, the relation
\begin{equation}
\mathcal{N}_{\tilde f(T)}(\mathcal N_f(T))=T
\end{equation}
holds for each face $f$ of $T$.

\begin{figure}
   \def\svgwidth{0.40\textwidth}
   \center
   \includegraphics{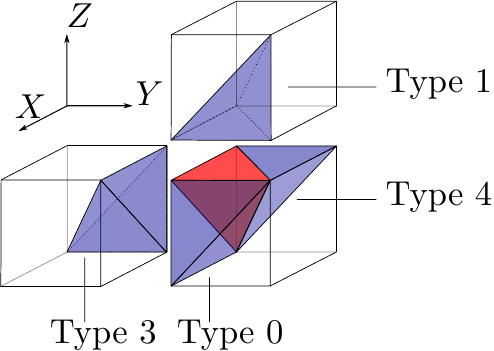}
    \caption{A tetrahedron $T$ of type $5$ (in the middle, red) and its four
face neighbors (blue) of types $1,0,4$, and $3$, drawn with their associated
cubes (exploded view). We see that the type of
$T$'s neighbors depends only on its type, while their node coordinates relative
to $T$'s depend additionally on $T$'s level.
(Color available online.)
}\label{fig:neighbor3}
\end{figure}

Our aim is to compute the Tet-id of $\mathcal{N}_f(T)$ and $\tilde{f}(T)$
from the Tet-id of $T$.
Using the underlying cube structure this problem can be solved for each occuring type of $d$-simplex separately,
and the solution scheme
is independent of the coordinates of $T$ and of $\ell$.
In Figure \ref{fig:neighbor3} the situation for a tetrahedron of type $5$ is illustrated, and in Tables \ref{tab:faceneighbor2d}
and \ref{tab:faceneighbor}
we present the general solution for each type.

Using these tables, a constant-time computation of the Tet-id of $\mathcal{N}_f(T)$ and of $\tilde{f}(T)$
from the Tet-id of $T$ is possible, and the 3D case
 is carried out in Algorithm~\ref{alg:face-neighbor}.
Note that this algorithm uses arithmetic expressions in $T.b$ to avoid the sixfold distinction of cases.

\begin{remark}
To find existing neighbors in a nonuniform refinement we use
Algorithm \ref{alg:face-neighbor} in combination with \texttt{Parent}
and \texttt{TM-Child} and comparison functions.

Of course it is possible that $\mathcal{N}_f(T)$ does not belong to $\mathcal T_d^\ell$ any longer.
If this is the case, then $f$ was part of the boundary of the root simplex $T^0_d$.
We describe in the next section how
we can decide in constant time whether $\mathcal{N}_f(T)$ is in $T^\ell_d$ or not.
\end{remark}

\begin{table}
\begin{center}
\begin{tabular}{|c|c|ccc|}
\hline
 \mytabvspace
 \mytabvspace
 2D & $f$              & 0 & 1 & 2   \\[0.2ex]\hhline{|=|=|===|}
\multirow{5}{*}{$T.b=0$} & \mytabvspace $\mathcal{N}_f.b$ & 1 & 1 & 1  \\[0.2ex]
& $\mathcal{N}_f.x$ & $T.x+h$ & $T.x$ & $T.x$\\[0.2ex]
& $\mathcal{N}_f.y$ & $T.y$ & $T.y$ & $T.y-h$\\[0.2ex]
& $\tilde{f}$      & 2 & 1 & 0       \\[0.2ex]\hline

\multirow{5}{*}{$T.b=1$} & \mytabvspace $\mathcal{N}_f.b$ & 0 & 0 & 0   \\[0.2ex]
 &$\mathcal{N}_f.x$ & $T.x$ & $T.x$ & $T.x-h$ \\[0.2ex]
 &$\mathcal{N}_f.y$ & $T.y+h$ & $T.y$ & $T.y$\\[0.2ex]
 &$\tilde{f}$      & 2 & 1 & 0       \\[0.2ex]\hline
\end{tabular}
\end{center}
\caption{Face neighbors in 2D.
      For each possible type $b\in\set{0,1}$ of a triangle $T$ and each of its
faces $f=f_i$, $i\in\set{0,1,2}$, the type, anchor node coordinates, and
corresponding face number $\tilde f$ of $T$'s neighbor across
      $f$ are shown.
      In the 2D case we can directly compute $\mathcal N.b = 1-T.b$ and $\tilde f = 2-f$.
      Here, $h=2^{\mathcal{L}-\ell}$ refers to the length of one square of level $\ell$.}
\label{tab:faceneighbor2d}
\end{table}

\begin{table}
\resizebox{0.5\textwidth}{!}{
\begin{tabular}{|c|c|cccl|}
\hline
 \mytabvspace
 \mytabvspace
 3D & $f$              & 0 & 1 & 2 & 3  \\[0.2ex]\hhline{|=|=|====|}
\multirow{5}{*}{$T.b=0$} & \mytabvspace $\mathcal{N}_f.b$ & 4 & 5 & 1 & 2  \\[0.2ex]
& $\mathcal{N}_f.x$ & $T.x+h$ & $T.x$ & $T.x$ & $T.x$\\[0.2ex]
& $\mathcal{N}_f.y$ & $T.y$ & $T.y$ & $T.y$ & $T.y-h$\\[0.2ex]
& $\mathcal{N}_f.z$ & $T.z$ & $T.z$ & $T.z$ & $T.z$\\[0.2ex]
& $\tilde{f}$      & 3 & 1 & 2 & 0       \\[0.2ex]\hline

\multirow{5}{*}{$T.b=1$} & \mytabvspace $\mathcal{N}_f.b$ & 3 & 2 & 0 & 5  \\[0.2ex]
 &$\mathcal{N}_f.x$ & $T.x+h$ & $T.x$ & $T.x$ & $T.x$\\[0.2ex]
 &$\mathcal{N}_f.y$ & $T.y$ & $T.y$ & $T.y$ & $T.y$\\[0.2ex]
 &$\mathcal{N}_f.z$ & $T.z$ & $T.z$ & $T.z$ & $T.z-h$\\[0.2ex]
 &$\tilde{f}$      & 3 & 1 & 2 & 0       \\[0.2ex]\hline

\multirow{5}{*}{$T.b=2$} & \mytabvspace$\mathcal{N}_f.b$ & 0 & 1 & 3 & 4  \\[0.2ex]
 &$\mathcal{N}_f.x$ & $T.x$ & $T.x$ & $T.x$ & $T.x$\\[0.2ex]
 &$\mathcal{N}_f.y$ & $T.y+h$ & $T.y$ & $T.y$ & $T.y$\\[0.2ex]
 &$\mathcal{N}_f.z$ & $T.z$ & $T.z$ & $T.z$ & $T.z-h$\\[0.2ex]
 &$\tilde{f}$      & 3 & 1 & 2 & 0       \\[0.2ex]\hline
 \end{tabular}
 }
 \hspace{-2ex}
\resizebox{0.5\textwidth}{!}{
 \begin{tabular}{|c|c|cccl|}
\hline
 \mytabvspace
 \mytabvspace
 3D & $f$              & 0 & 1 & 2 & 3  \\[0.2ex]\hhline{|=|=|====|}
\multirow{5}{*}{$T.b=3$} & \mytabvspace$\mathcal{N}_f.b$ & 5 & 4 & 2 & 1  \\[0.2ex]
 &$\mathcal{N}_f.x$ & $T.x$ & $T.x$ & $T.x$ & $T.x-h$\\[0.2ex]
 &$\mathcal{N}_f.y$ & $T.y+h$ & $T.y$ & $T.y$ & $T.y$\\[0.2ex]
 &$\mathcal{N}_f.z$ & $T.z$ & $T.z$ & $T.z$ & $T.z$\\[0.2ex]
 &$\tilde{f}$      & 3 & 1 & 2 & 0       \\[0.2ex]\hline

\multirow{5}{*}{$T.b=4$} & \mytabvspace$\mathcal{N}_f.b$ & 2 & 3 & 5 & 0  \\[0.2ex]
 &$\mathcal{N}_f.x$ & $T.x$ & $T.x$ & $T.x$ & $T.x-h$\\[0.2ex]
 &$\mathcal{N}_f.y$ & $T.y$ & $T.y$ & $T.y$ & $T.y$\\[0.2ex]
 &$\mathcal{N}_f.z$ & $T.z+h$ & $T.z$ & $T.z$ & $T.z$\\[0.2ex]
 &$\tilde{f}$      & 3 & 1 & 2 & 0       \\[0.2ex]\hline

\multirow{5}{*}{$T.b=5$} & \mytabvspace$\mathcal{N}_f.b$ & 1 & 0 & 4 & 3  \\[0.2ex]
 &$\mathcal{N}_f.x$ & $T.x$ & $T.x$ & $T.x$ & $T.x$\\[0.2ex]
 &$\mathcal{N}_f.y$ & $T.y$ & $T.y$ & $T.y$ & $T.y-h$\\[0.2ex]
 &$\mathcal{N}_f.z$ & $T.z+h$ & $T.z$ & $T.z$ & $T.z$\\[0.2ex]
 &$\tilde{f}$      & 3 & 1 & 2 & 0       \\[0.2ex]\hline
\end{tabular}
}
\caption{Face neighbors in 3D.
For each possible type $b\in\set{0,1,2,3,4,5}$ of a tetrahedron $T$
and each of its faces $f=f_i$, $i \in\set{0,1,2,3}$
the type  $\mathcal{N}_f(T).b$ of $T$'s neighbor across $f$, its coordinates of the
anchor node $\mathcal{N}_f(T).x$, $\mathcal{N}_f(T).y$, $\mathcal{N}_f(T).z$
and the corresponding face number $\tilde{f}(T)$, across which $T$ is
$\mathcal{N}_f(T)$'s neighbor, are shown.} 
\label{tab:faceneighbor}
\end{table}

\begin{algorithm}
 \caption{\texttt{Face-neighbor}(\aTet{} $T$,\Int $f$)}
 \label{alg:face-neighbor}
 \DontPrintSemicolon
 $b\gets T.b$,
 $x_0\gets T.x_0$, $x_1\gets T.x_1$, $x_2\gets T.x_2$\;
 \eIf{$f=1$ or $f=2$}{
    $\tilde{f}\gets f$
    \eIf{$(b\algomod2=0 \textnormal{\algoand} f=2)\textnormal\algor(b\algomod2\neq0 \textnormal\algoand f=1)$}{
      $b\gets b+1$
    }
    {
      $b\gets b-1$
    }
 }
 {%
  $\tilde{f}\gets 3-f$

  $h\gets 2^{\mathcal{L}-T.\ell}$

  \eIf(\IfComment{$f=0$}){$f=0$}{%
    $i\gets b\textnormal{\textbf{ div }}2$\;
    $x_i\gets T.x_i+h$\;
    $b\gets b + (b\algomod2=1\abst{?}2:4)$\;
  }%
  (\IfComment{$f=3$}){%
    $i\gets (b+3)\algomod6 \textnormal{\textbf{ div }}2$\;
    $x_i \gets T.x_i - h$\;
    $b\gets b + (b\algomod2=0\abst{?}2:4)$\;
  }%
  }%
\Return $(x_0,x_1,x_2,b\algomod 6,\tilde{f})$\;
\end{algorithm}

For completeness, we summarize the geometry and numbers of
$d$-simplices tou\-ching each other via a corner ($d=2$ or $d=3$) or edge (only
$d=3$).
In this paper we do not list these neighboring tetrahedra in detail.

For $d=3$ each corner in the mesh $\mathcal T^\ell_3$ has 24 adjacent tetrahedra;
thus each tetrahedron has at each corner 23 other tetrahedra that share
this particular corner.
For $d=2$ the situation is similar, with six triangles meeting at each corner.
To examine the number of adjacent tetrahedra to an edge
we distinguish three types of edges in $\mathcal T^\ell_d$:
\begin{enumerate}
 \item edges that are also edges in the underlying hexahedral mesh;
 \item edges that are the diagonal of a side of a cube in the hexahedral mesh;
 \item edges that correspond to the inner diagonal of a cube in the hexahedral mesh.
\end{enumerate}
Edges of the first and third kind have six adjacent tetrahedra each,
and edges of the second kind do have four adjacent tetrahedra each.

\subsection{The exterior of the root simplex}
\label{sec:outside}
When computing neighboring $d$-simplices it is possible that the neighbor in question does not belong to the root simplex
$T^0_d$ but lies outside of it.
If we look at face-neighbors of a $d$-sim\-plex $T$, the fact that the considered neighbor lies outside
means that the respective face was on the boundary of $T^0_d$.
In order to check whether a computed $d$-simplex is outside the base simplex, we investigate
a more general problem:
Given anchor node coordinates $(x_0,y_0)^T\in\IZ^2$, respectively $(x_0,y_0,z_0)^T\in\IZ^3$, of type $b$ a level $\ell$,
decide whether the corresponding $d$-simplex $N$ lies inside or outside
of the root tetrahedron $T^0_d$: $N\in\mathcal{T}^\ell_d$ or $N\notin\mathcal{T}^\ell_d$.
At the end of this section we furthermore generalize this to the problem of deciding for any two $d$-simplices $N$ and $T$
whether or not $N$ lies outside of $T$.
We solve this problem in constant time and independent of the levels of $N$ and $T$.

We examine the 3D case.
Looking at $T^0_3$ we observe that two of its boundary faces correspond to faces of the root cube,
namely, the intersections of $T^0_3$ with the $y=0$ and the $x=2^\mathcal{L}$ planes.
The other two boundary faces of $T^0_3$ are the intersections with the $x=z$ and the $y=z$ planes.
Thus, the boundary of $T^0_3$ can be described as the intersection of $T^0_3$ with those planes.
We refer to the latter two planes as $E_1$ and $E_2$.

\begin{SCfigure}
   \def\svgwidth{46ex}
   \includegraphics{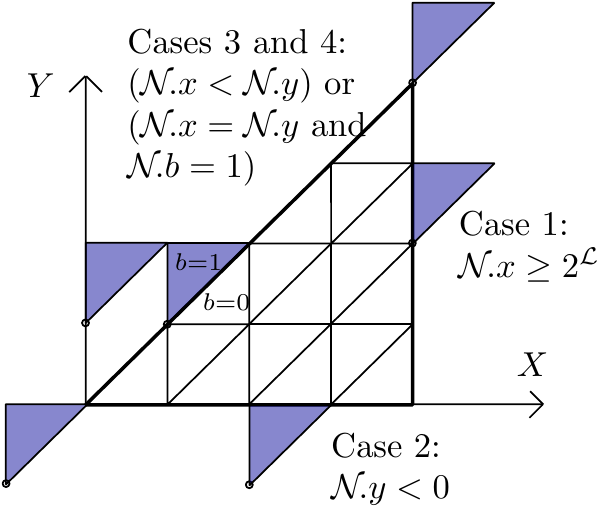}
   \caption{A uniform level 2 refinement of the triangle $T^0$ in 2D and triangles lying outside of it with their anchor nodes marked.
          When deciding whether a triangle with given anchor node coordinates lies outside of $T^0$ there
          are four cases to consider, one for each face of $T^0$.
          For the two faces lying parallel to the $X$-axis, respectively, $Y$-axis, it suffices to check whether
          the $x$-coordinate is greater than or equal to $2^\mathcal{L}$, or whether the $y$-coordinate is smaller than $0$.
          Similarly one can conclude that the triangle lies outside of $T^0$ if its $x$-coordinate is smaller than
          its $y$-coordinate. If both coordinates agree (and none of the previous cases applies) then
          the given triangle is outside $T^0$ if and only if its type is $1$.}
   \label{fig:neighboroutside2d}
\end{SCfigure}

Let $N$ be a tetrahedron with anchor node $(x_0,y_0,z_0)^T\in\IZ^3$ of type $b$ and level $\ell$
and denote with $(x_i,y_i,z_i)^T$ the coordinates of node $i$ of $N$.
Since $(x_i,y_i,z_i)^T\geq(x_0,y_0,z_0)^T$
(componentwise), we directly conclude that if $x_0\geq 2^\mathcal L$ or $y_0<0$ then $N\notin\mathcal T_3$.
Because the outer normal vectors of $T^0_3$ on the two faces intersecting with $E_1$ and $E_2$ are
\begin{equation}
\vec{n}_1= \frac{1}{\sqrt 2}\begin{pmatrix}
  -1\\0\\1
 \end{pmatrix} \quad\textrm{and}\quad
\vec{n}_2= \frac{1}{\sqrt 2}\begin{pmatrix}
  0\\1\\-1
 \end{pmatrix},
\end{equation}
we also conclude that $N\notin\mathcal T_3$ if $z_0-x_0>0$ or $y_0-z_0>0$.
Now we have already covered all the cases except those where the anchor node of $N$ lies directly in $E_1$ or $E_2$.
In these cases we cannot solve the problem by looking at the coordinates of the anchor node alone, since there
exist tetrahedra $T'\in\mathcal{T}_3$ with anchor nodes lying in one of these planes
(see Figure \ref{fig:neighboroutside2d} for an illustration of the analogous case in 2D).
This depends on the type of the tetrahedron in question.
We observe that a tetrahedron $T'\in\mathcal{T}_3$ with anchor node lying in $E_1$ can have the types $0$ ,$1$, or $2$, and a tetrahedron
with anchor node lying in $E_2$ can have the types $0$, $4$, or $5$.
We conclude that to check whether $N$ is outside of the root tetrahedron we have to check
if any one of six conditions is fulfilled.
In fact these conditions fit into the general form below with $x_i = x$, $x_j=y$, $x_k=z$, and $T$ as the root tetrahedron;
thus $T.x = T.y = T.z = 0$ and $T.b=0$.

These generalized conditions solve the problem to check for any two given tetrahedra $N$ and $T$, whether
$N$ lies outside of $T$ or not.
\begin{proposition}
\label{prop:neighboroutside}
 Given two $d$-simplices $N, T$ with $N.\ell> T.\ell$, then $N$ is outside
of the simplex $T$---which is equivalent to saying that $N$ is no descendant of
$T$---if and only if at least one of the following conditions is fulfilled.

 For 2D,
\begin{subequations}
\begin{align}
&N.x_i - T.x_i \geq 2^{\mathcal L - T.\ell},\hphantom{xxxxxxxxxxxxxxxxxxxxxxxxxxxxxxxxxxxxxxxxxx}\\
&N.x_j-T.x_j<0,\\
&(N.x_j-T.x_j)-(N.x_i-T.x_i)>0,\\
&N.x_i-T.x_i=N.x_j-T.x_j\textrm{ and }N.b=\left\lbrace\begin{array}{cc}
                                             1 &\textrm{if\, } T.b = 0, \\
                                             0 &\textrm{if\, } T.b = 1.
                                            \end{array}\right.
\end{align}
\end{subequations}
For 3D,
\begin{subequations}
\begin{align}
 &N.x_i -T.x_i\geq 2^{\mathcal L - T.\ell},\hphantom{xxxxxxxxxxxxxxxxxxxxxxxxxxxxxxxxxxxxxxxxxx}\\
 &N.x_j-T.x_j<0,\\
 &(N.x_k-T.x_k)-(N.x_i-T.x_i)>0,\\
 &(N.x_j-T.x_j)-(N.x_k-T.x_k)>0,\\[1ex]
 &\begin{aligned}
 &N.x_j-T.x_j=N.x_k-T.x_k \\ &\mathrm{ and }\quad N.b\in\left\lbrace\begin{array}{cc}
                                             \set{T.b+1,T.b+2,T.b+3}, & \textrm{if\, } T.b \textrm{ is even,} \\
                                             \set{T.b-1,T.b-2,T.b-3}, & \textrm{if\, }T.b \textrm{ is odd, }
                                            \end{array}\right.\\
 \end{aligned}\\
 &%
 \begin{aligned}
 &N.x_k-T.x_k=N.x_i-T.x_i\\ &\mathrm{ and }\quad N.b\in\left\lbrace\begin{array}{cc}
                                             \set{T.b-1,T.b-2,T.b-3}, &\textrm{if\, } T.b\textrm{ is even,} \\
                                             \set{T.b+1,T.b+2,T.b+3}, &\textrm{if\, } T.b\textrm{ is odd. }
                                            \end{array}\right.
 \end{aligned} \\
&%
\begin{aligned}
 &N.x_j - T.x_j = N.x_k - T.x_k \quad\mathrm { and }\quad N.x_i - T.x_i = N.x_k - T.x_k\quad
\mathrm{ and }\quad N.b \neq T.b
\end{aligned}
\end{align}
\end{subequations}
 The coordinates $x_i$, $x_j$, and $x_k$ are chosen in dependence of the type of $T$ according to Table \ref{table:xixjxk}.
\begin{table}
\centering
\raisebox{1.3ex}{
\begin{tabular}{|c|cc|}
 \hline
  & \multicolumn{2}{c|}{$T.b$} \\
  $2$D
        & 0 & 1 \\ \hline
  $x_i$ & $x$ & $y$  \\
  $x_j$ & $y$ & $x$  \\ \hline
 \end{tabular}}
 \begin{tabular}{|c|cccccc|}
 \hline
  & \multicolumn{6}{c|}{$T.b$} \\
   $3$D   & 0 & 1 & 2 & 3 & 4 & 5 \\ \hline
  $x_i$ & $x$ & $x$ & $y$ & $y$ & $z$ & $z$ \\
  $x_j$ & $y$ & $z$ & $z$ & $x$ & $x$ & $y$ \\
  $x_k$ & $z$ & $y$ & $x$ & $z$ & $y$ & $x$ \\
  \hline
 \end{tabular}
\caption{Following the general scheme described in this section to compute whether a given $d$-simplex $N$ lies outside
        of another given $d$-simplex $T$, we give the coordinates $x_i$, $x_j$, and $x_k$ in dependence of the type of $T$.}
\label{table:xixjxk}
\end{table}
\begin{proof}
By shifting $N$ by $(-T.\vec x)$ we reduce the problem to checking whether the shifted $d$-simplex lies outside of a
$d$-simplex with anchor node $\vec 0$, level $T.\ell$ and type $T.b$.
For $d=3$ the proof is analogous to the above argument, where we considered the case $b=0$ and $\ell = 0$.
In two dimensions the situation is even simpler, since there exists only one face of the root
triangle that is not a coordinate axis (see Figure \ref{fig:neighboroutside2d}).
\end{proof}
\end{proposition}

\subsection[A consecutive index]{A consecutive index}
\label{sec:consecindex}

In contrast to the Morton index for cubes, the TM-index for $d$-simplices does not
produce a consecutive range of numbers.
Therefore, two simplices $T$ and $T'$ of level $\ell$
that are direct successors/predecessors with respect to the tetrahedral Morton order do not
necessarily fulfill $m(T)=m(T')\pm 2^{d(\mathcal L-\ell)}$ or $m(T)=m(T')\pm 1$.
For $d=3$ this follows directly from the fact that each $b^j$ occupies three bits,
but there are only six values that each $b^j$ can assume, since there are only six different types.
In 2D this follows from the fact that not every combination of anchor
node coordinates and type can occur for triangles in $\mathcal T_2$,
the triangle with anchor node $(0,0)$ and type $1$ being one example.
This also means that the largest occurring TM-index is bigger than $2^{d\mathcal{L}}-1$.

Constructing a consecutive index that respects the order given by the TM-index
is possible, as we show in this section.
Since in practice it is more convenient to work with this consecutive index instead
of the TM-index,
our aim is to construct for each level $\ell$ a consecutive index
$
I(T)\in\set{0,\dots,2^{d\ell}-1},
$
such that
\begin{equation}
 \forall\ T,S\in\mathcal{T}^\ell_d\colon\quad I(T)<I(S) \abst\Leftrightarrow m(T)<m(S).
\end{equation}
This index can also be understood as a bijection
\begin{equation}
  I\colon \set{m\in\IN_0 \abst | m = m(T) \textnormal{ for a } T\in\mathcal T^\ell_d}
  \xrightarrow{\cong}
  \set{0,\ldots,2^{d\ell}-1},
\end{equation}
mapping the TM-indices of level $\ell$ $d$-simplices to a consecutive range of numbers.
It is obvious that $I(T^0_d)=0$.
The index $I(T)$ can be easily
computed as the $\ell$-digit $2^d$-ary number consisting of the local indices as digits, thus
\begin{equation}
\label{eq:gloindex}
 I(T) = (I_\mathrm{loc}(T^1),\dots,I_\mathrm{loc}(T^\ell))_{2^d}.
\end{equation}

Algorithm \ref{alg:ComputeIndexfromtet} shows an implementation of this computation.
It can be done directly from the Tet-id of $T$, and thus it is not
necessary to compute the TM-index of $T$ first.

\begin{algorithm}
 \caption{$I$(\aTet{}$T$)}
 \label{alg:ComputeIndexfromtet}
 \DontPrintSemicolon
 $I \gets 0$,
 $b \gets T.b$\;
 \For{$T.\ell \geq i \geq 1$}{
    $c \gets \texttt{c-id}(T,i)$\;
    $I\gets I + 8^i I_\mathrm{loc}^{b}(c)$
    \Comment{See Table \ref{table:candbtoiloc}; multiply with $4^i$ for 2D}
    $ b \gets \textrm{Pt}(c, b)$\;
 }
 \Return $I$\;
\end{algorithm}

The inverse operation of computing $T$ from $I(T)$ and a given level $\ell$ can be carried out in a similar fashion;
see Algorithm \ref{alg:Init_id}.
For each $0\leq i\leq\ell$ we look up the type $b$ and the cube-id of $T^i$
from $I_\mathrm{loc}(T^{i})$ and the type of $\texttt{Parent}(T^i)=T^{i-1}$ (starting with $\type(T^0)=0$) via Tables \ref{table:IlocPbtocubeid} and \ref{table:IlocPbtotype}.
From the cube-ids we can build up the anchor node coordinates of $T$. The last computed
type is the type of $T$.
The runtime of this algorithm is $\mathcal O(T.\ell)$.

\begin{algorithm}
 \caption{\texttt{T}(consecutive index $I$,\Int $\ell$)}
 \label{alg:Init_id}
 \DontPrintSemicolon
 $T.x,T.y,T.z\gets 0$,
 $b \gets 0$\;
 \For{$1\leq i \leq \ell $}{
    Get $I_\mathrm{loc}(T^i)$ from $I$\Comment{See \eqref{eq:gloindex}}
    $c\gets\texttt{c-id}(T^i),\, b\gets T^i.b$\Comment{See Tables \ref{table:IlocPbtocubeid} and \ref{table:IlocPbtotype}}
    \lIf{$c\bitwand 1$}{$T.x\gets T.x + 2^{\mathcal L -i}$}
    \lIf{$c\bitwand 2$}{$T.y\gets T.y + 2^{\mathcal L -i}$}
    \lIf(\IfComment{Remove this line for 2D}){$c\bitwand 4$}{$T.z\gets T.z + 2^{\mathcal L -i}$}
 }
 $T.b\gets b$\;
 \Return $T$\;
\end{algorithm}

Similar to Algorithm \ref{alg:ComputeIndexfromtet} is Algorithm
\texttt{Is\_valid}, which decides whether a given index $m\in[0,2^{6\mathcal{L}})
\cap \IZ$ is in fact a TM-index for a tetrahedron.
Thus, in the spirit of Section \ref{sec:Prop} we can decide whether a given
6D cube is in the image of map (\ref{eq:MapTQ}) that embeds
$\mathcal{T}_3$ into the set of 6D subcubes of $[0,2^{\mathcal{L}}]^6$.
The runtime of \texttt{Is\_valid} is $\mathcal{O}(\mathcal{L})$.

\begin{algorithm}
 \caption{\texttt{Is\_valid}($m\in [0,2^{6\mathcal{L}}) \cap \IZ, \ell$)}
 \label{alg:IsValid}
 \DontPrintSemicolon
 $I \gets 0$\;
 $k \gets 6(\mathcal{L}-i)$\;
 \For{$\ell \geq i \geq 1$}{
    $b \gets (m_{k},m_{k + 1},m_{k + 2})_8$\;
    $c \gets (m_{k + 3},m_{k+4},m_{k+5})_8$\;
    $k \gets 6(\mathcal{L}-i + 1)$\;
    \If (\IfComment {Take $(0,0,0)_8$ if $i = 1$})
    {$(m_{k},m_{k + 1},m_{k + 2})_8 \neq \textrm{Pt}(c, b)$}
    {\Return \texttt{False}}}
 \Return \texttt{True}\;
\end{algorithm}

\begin{table}
\begin{center}
\raisebox{5.9ex}{
\begin{tabular}{|rc|cccl|}
\hline
\multicolumn{2}{|c|}{{$I^b_\mathrm{loc}(c)$}}&\multicolumn{4}{c|}{cube-id c}\\
 \multicolumn{2}{|c|}{2D} &\mytabvspace
 $0$ & $1$ & $2$ & $3$ \\[0.5ex]\hline
 \multirow{2}{*}{b}&\mytabvspace  0&  0 & 1 & 1 & 3 \\[0.2ex]
  &1&  0 & 2 & 2 & 3\\ \hline
\end{tabular}
}
\begin{tabular}{|rc|cccccccl|}
\hline
\multicolumn{2}{|c|}{{$I^b_\mathrm{loc}(c)$}}&\multicolumn{8}{c|}{cube-id c}\\
 \multicolumn{2}{|c|}{3D} &\mytabvspace
 $0$ & $1$ & $2$ & $3$ & $4$ & $5$ & $6$ & $7$\\[0.5ex]\hline
 \multirow{6}{*}{b}&\mytabvspace
  0&0& 1& 1& 4& 1& 4& 4& 7\\[0.2ex]
 &1&0& 1& 2& 5& 2& 5& 4& 7\\[0.2ex]
 &2&0& 2& 3& 4& 1& 6& 5& 7\\[0.2ex]
 &3&0& 3& 1& 5& 2& 4& 6& 7\\[0.2ex]
 &4&0& 2& 2& 6& 3& 5& 5& 7\\[0.2ex]
 &5&0& 3& 3& 6& 3& 6& 6& 7\\  \hline
\end{tabular}
\caption{The local index of a tetrahedron $T\abst{\in}\mathcal{T}$ in dependence of its cube-id $c$ and type $b$.}
\label{table:candbtoiloc}
\end{center}
\end{table}

\begin{table}
\begin{center}
\raisebox{5.9ex}{
\begin{tabular}{|rc|cccl|}
\hline
\multicolumn{2}{|c|}{\mytabvspace {$\cid(T)$}}&\multicolumn{4}{c|}{$I_\mathrm{loc}(T)$}\\
\multicolumn{2}{|c|}{2D} &
 \mytabvspace
 $0$ & $1$ & $2$ & $3$ \\[0.5ex]\hline
 \multirow{2}{*}{$P.b$}&\mytabvspace0& 0 & 1  & 1 &  3 \\[0.2ex]
 &1& 0 & 2 & 2 & 3 \\ \hline
\end{tabular}\hspace{0.5ex}
}%
\begin{tabular}{|rc|cccccccl|}
\hline
\multicolumn{2}{|c|}{{\mytabvspace $\cid(T)$}}&\multicolumn{8}{c|}{$I_\mathrm{loc}(T)$}\\
\multicolumn{2}{|c|}{3D}&
\mytabvspace $0$ & $1$ & $2$ & $3$ & $4$ & $5$ & $6$ & $7$\\[0.5ex]\hline
 \multirow{6}{*}{$P.b$}&\mytabvspace0& 0 & 1 & 1 & 1 & 5 & 5 & 5 & 7  \\[0.2ex]
 &1& 0 & 1 & 1 & 1 & 3 & 3 & 3 & 7     \\[0.2ex]
 &2& 0 & 2 & 2 & 2 & 3 & 3 & 3 & 7    \\[0.2ex]
 &3& 0 & 2 & 2 & 2 & 6 & 6 & 6 & 7    \\[0.2ex]
 &4& 0 & 4 & 4 & 4 & 6 & 6 & 6 & 7    \\[0.2ex]
 &5& 0 & 4 & 4 & 4 & 5 & 5 & 5 & 7    \\ \hline
\end{tabular}
\caption{For a tetrahedron $T\abst{\in}\mathcal{T}$ of local Index $I_\mathrm{loc}$ whose parent $P$ has type $P.b$ we give
       the cube-id of $T$.}
\label{table:IlocPbtocubeid}
\end{center}
\end{table}

\begin{table}
\begin{center}
\raisebox{5.9ex}{
\begin{tabular}{|rc|cccl|}
\hline
\multicolumn{2}{|c|}{\mytabvspace {$T.b$}}&\multicolumn{4}{c|}{$I_\mathrm{loc}(T)$}\\
\multicolumn{2}{|c|}{2D} &
 \mytabvspace
 $0$ & $1$ & $2$ & $3$ \\[0.5ex]\hline
 \multirow{2}{*}{$P.b$}&\mytabvspace0& 0 & 0 & 1 & 0 \\[0.2ex]
 &1& 1 & 0 & 1 & 1  \\ \hline
\end{tabular}\hspace{0.5ex}
}%
\begin{tabular}{|rc|cccccccl|}
\hline
\multicolumn{2}{|c|}{{\mytabvspace $T.b$}}&\multicolumn{8}{c|}{$I_\mathrm{loc}(T)$}\\
\multicolumn{2}{|c|}{3D}&
\mytabvspace $0$ & $1$ & $2$ & $3$ & $4$ & $5$ & $6$ & $7$\\[0.5ex]\hline
 \multirow{6}{*}{$P.b$}&\mytabvspace0& 0 & 0  & 4 & 5 & 0 & 1 & 2 & 0  \\[0.2ex]
 &1& 1 & 1 & 2 & 3 & 0 & 1 & 5 & 1     \\[0.2ex]
 &2& 2 & 0 & 1 & 2 & 2 & 3 & 4 & 2    \\[0.2ex]
 &3& 3 & 3 & 4 & 5 & 1 & 2 & 3 & 3    \\[0.2ex]
 &4& 4 & 2 & 3 & 4 & 0 & 4 & 5 & 4    \\[0.2ex]
 &5& 5 & 0 & 1 & 5 & 3 & 4 & 5 & 5    \\ \hline
\end{tabular}
\caption{For a tetrahedron $T\abst{\in}\mathcal{T}$ of local Index $I_\mathrm{loc}$ whose parent $P$ has type $P.b$ we give
       the type of $T$.}
\label{table:IlocPbtotype}
\end{center}
\end{table}

The consecutive index simplifies the relation between the TM-index of a simplex
and its position in the SFC.
In the special case of a uniform mesh, the consecutive index and the position
are identical.

\subsection{Successor and predecessor}
Calculating the TM-index corresponding to a particular consecutive index is
occasionally needed in higher-level algorithms.
This is relatively expensive,
since it involves a loop over all refinement levels, thus some 10 to 30 in extreme cases.
However often the task is to compute a whole range of $d$-simplices.
This occurs, for example, when creating an initial uniform refinement of a given mesh
(see Algorithm \texttt{New} in Section \ref{sec:new}).
That is, for a given consecutive index $I$, a level $\ell$, and a
count $n$, find the $n$ level-$\ell$ simplices following the $d$-simplex corresponding to the consecutive index $I$,
that is, the $d$-simplices corresponding to the $n$ consecutive indices
$I,I+1,\dots,I+n-1$.
Ideally, this operation should run linearly in $n$, independent of $\ell$, but if we used Algorithm \ref{alg:Init_id} to create each of the $n+1$ simplices
we would have a runtime of $\mathcal{O}(n\mathcal{L})$.
In order to achieve the desired linear runtime
we introduce the operations \texttt{Successor} and \texttt {Predecessor} that run in average $\mathcal O(1)$ time.
These o\-pera\-tions compute from a given $d$-simplex $T$ of level $\ell$ with consecutive index $I_T$ the $d$-simplex $T'$ whose
consecutive index is $I_T+1$, respectively, $I_T-1$.
Thus, $T'$ is the next level $\ell$ simplex in the SFC after $T$ (resp.\ the previous one).
Algorithm \ref{alg:successor}, which we introduce to solve this problem does not require knowledge about the consecutive indices
$I_T$ and $I_T\pm 1$ and can be computed significantly faster than Algorithm \ref{alg:Init_id}; see Lemma \ref{lem:IncmortonRuntime}.

\let\oldnl\nl
\newcommand{\nonl}{\renewcommand{\nl}{\let\nl\oldnl}}
\begin{algorithm}
\DontPrintSemicolon
\SetKwFunction{proc}{proc}
\caption{\texttt{Successor}(\aTet{} $T$)}
\label{alg:successor}
\Return \texttt{Successor\_recursion}$(T,T,T.\ell)$\\[2ex]

\setcounter{AlgoLine}{0}
\nonl \textbf{Function} \texttt{Successor\_recursion}(\Tet{} $T$,\Tet{} $T'$,\Int $\ell$)\;
$c\gets$ \texttt{c-id}$(T,\ell)$\;
From $c$ and $b$ look up $i:=I_\mathrm{loc}(T^\ell)$\Comment{See Table \ref{table:candbtoiloc}}
$i\gets (i+1)\abst \algomod 8$\label{line:changeforpred}\;
\eIf(\IfComment{Enter recursion (in rare cases)}){$i = 0$}{
  $T' \gets$ \texttt{Successor\_recursion} $(T,T',\ell-1)$\;
  $\hat b\gets T'.b$\Comment{$\hat b$ stores the type of $T'^{\ell-1}$}
}{
$\hat b\gets$ Pt$(c,b)$\;
}
From $\hat b$ and $I_\mathrm{loc} = i$ look up $(c',b')$\Comment{See Tables \ref{table:IlocPbtocubeid} and \ref{table:IlocPbtotype}}
Set the level $\ell$ entries of $T'.x, T'.y$ and $T'.z$ to $c'$\;
$T'.b\gets b'$\;
\Return $T'$\;
\end{algorithm}
To compute the predecessor of $T$ we only need to reverse the sign in Line \ref{line:changeforpred} in the
\texttt{Successor\_re\-cur\-sion} subroutine of Algorithm \ref{alg:successor}.

\begin{lemma}
\label{lem:IncmortonRuntime}
 Algorithm \ref{alg:successor} has constant average runtime (independent of $\mathcal{L}$).
 \begin{proof}
  Because each operation in the algorithm can be executed in constant time, the average runtime is $nc$, where $c$ is a
  constant (independent of $\mathcal{L}$) and $n-1$ is the number of average recursion steps.
  Since in consecutive calls to the algorithm the variable $i$ cycles through
$0$ to $2^d-1$ we conclude that the recursion is on average executed in every
$2^d$th step, allowing for a geometric series argument.
 \end{proof}
\end{lemma}

We see in Algorithm \ref{alg:successor} the usefulness of the consecutive index.
Because we are using this index instead of the TM-index, computing
the index of the successor/pre\-de\-cessor only requires adding/subtracting $1$ to the given index.
On the other hand, computing the TM-index of a successor/predecessor would involve more subtle computations.
\pagebreak

\section{High-level AMR algorithms}

To develop the complete AMR functionality required by numerical applications,
we aim at a forest of quad-/octrees in the spirit of
\cite{BursteddeWilcoxGhattas11, IsaacBursteddeWilcoxEtAl15}.
Key top-level algorithms are:
\begin{itemize}
 \item \texttt{New}.
   Given an input mesh of conforming simplices, each considered a root simplex,
   generate an initial refinement.
 \item \texttt{Adapt}. Adapt (refine and coarsen) a mesh according to a given criterion.
 \item \texttt{Partition}. Partition a mesh among all processes such that the load
is balanced, possibly according to weights.
 \item \texttt{Balance}. Establish a 2:1 size condition between neighbors in a given refined mesh.
              The levels of any two neighboring simplices must differ by at most 1.
 \item \texttt{Ghost}. For each process, assemble the layer of directly
     neighboring elements owned by other processes.
 \item \texttt{Iterate}. Iterate through the local mesh, executing a callback
     function on each element and on all interelement interfaces.
\end{itemize}
Since partitioning via SFC only uses the SFC index as information, we refer to already existing descriptions of
\texttt{Partition} for hexahedral or simplicial SFCs \cite{PilkingtonBaden94,BursteddeWilcoxGhattas11}.
\texttt{Balance}, \texttt{Ghost}, and \texttt{Iterate} are sophisticated parallel algorithms and require
additional theoretical work, which is beyond the scope of this paper.

Here, we briefly describe \texttt{New} and \texttt{Adapt}.
In the forest-of-trees approach we model an adaptive mesh
by a coarse mesh of level $0$ $d$-simplices, the \emph{trees}.
Such a coarse mesh could be specified manually for simple geometries,
or obtained from executing a mesh generator.
Each level $0$ simplex is identified with the root simplex $T_d^0$
and then refined adaptively to produce the fine and potentially nonconforming
mesh of $d$-simplices.
These simplices are partitioned among all processes; thus each process holds
a range of trees, of which the first and last may be incomplete: Their leaves
are divided between multiple processes.

An entity $\mathcal F$ of the structure \texttt{forest} consists of  the following entries
\begin{itemize}
 \item $\mathcal{F}.C$ --- the coarse mesh;
 \item $\mathcal{F}.\mathcal{K}$ --- the process-local trees;
 \item $\mathcal{F}.\mathcal{E}_k$ --- for each local tree $k$ the list of process-local
                    simplices in tetrahedral Morton order.
\end{itemize}
We acknowledge that \texttt{New} and \texttt{Adapt} are essentially
communication-free, but still serve well to exercise some of the fundamental
algorithms described earlier.

\subsection{New}
\label{sec:new}

The \texttt{New} algorithm creates a partitioned uniform level $\ell$ refined forest from a given coarse mesh.
To achieve this, we first compute the first and last $d$-simplices belonging to the current process $p$.
From this range we can calculate which trees belong to $p$ and for each of these trees, the consecutive index of
the first and last $d$-simplices on this tree.
We then create the first simplex in a tree by a call to \texttt{T} (Algorithm \ref{alg:Init_id}).
In contrast to the \texttt{New} algorithm in \cite{BursteddeWilcoxGhattas11}
we create the remaining simplices by calls to \texttt{Successor} instead of \texttt{T} to
avoid the $\mathcal{O}(\ell)$ runtime of \texttt{T} in the case of simplices.
Our numerical tests, displayed in Figure \ref{fig:scalenew},
show that the runtime of \texttt{New} is in fact linear in the number of
elements and does not depend on the level $\ell$.
Within the algorithm, $K$ denotes the number of trees in the coarse mesh and 
$P$ the number of processes.
\begin{algorithm}
\caption{\texttt{New}(\texttt{Coarse Mesh} $C$, \Int{} $\ell$)}
\label{alg:New}
$n \gets 2^{d\ell}$, $N\gets nK$ \Comment{$d$-simplices per tree and global number of $d$-simplices}\vspace{0.8ex}
$g_\mathrm{first} \gets \left\lfloor {Np}/{P}\right\rfloor$,
$g_\mathrm{last} \gets \left\lfloor{N(p+1)}/{P}\right\rfloor-1$ \Comment{Global numbers of first and last..}\vspace{0.8ex}
$k_\mathrm{first} \gets \left\lfloor g_\mathrm{first}/n \right\rfloor,
\, k_\mathrm{last} \gets \left\lfloor g_\mathrm{last}/n \right\rfloor$\Comment{..local simplex and local tree range}\vspace{0.5ex}
\For{$t \in \set{k_\mathrm{first},\dots,k_\mathrm{last}}$}
{
  $e_\mathrm{first} \gets (t = k_\mathrm{first}) \abst ? g_\mathrm{first}-nt : 0$\;
  $e_\mathrm{last} \gets (t = k_\mathrm{last}) \abst ? g_\mathrm{last}-nt : n-1$\;
  $T\gets$\texttt{T} $(e_\mathrm{first},\ell)$\Comment{Call Algorithm \ref{alg:Init_id}}
  $\mathcal{E}_k \gets \set T$\;
  \For{$e \in \set{e_\mathrm{first},\dots,e_{last}-1}$}
  {
    $T\gets$ \texttt{Successor} $(T)$\;
    $\mathcal{E}_k \gets  \mathcal{E}_k \cup \set T$
  }
}
\end{algorithm}

After \texttt{New} returns, the process local number of elements is known, and
per-element data can be allocated linearly in an array of structures, or a
structure of arrays, depending on the specifics of the application.

\subsection{Adapt}

The \texttt{Adapt} algorithm modifies an existing forest by refining and
coar\-se\-ning the $d$-simplices of a given forest according to a callback
function.
It does this by traversing the $d$-simplices of each tree in tetrahedral Morton order
and passing them to the callback function. If the current $d$-simplex and its $2^d-1$ successors form a family (all having the same parent),
then the whole family is passed to the callback.
This callback function accepts either one or $2^d$ $d$-simplices as input
plus the index of the current tree.
In both cases, a return value greater than zero means that the first input $d$-simplex should be refined, and
thus its $2^d$ children are added in tetrahedral Morton order to the new
forest.  Additionally, if the input consists of $2^d$ simplices, they form a
family, and a return value
smaller than zero means that this family should be coarsened, thus replaced by their parent.
If the callback function returns zero, the first given $d$-simplex remains unchanged
and is added to the new forest, and \texttt{Adapt} continues
with the next $d$-simplex in the current tree.
The \texttt{Adapt} algorithm creates a new forest from the given one and can handle recursive refinement/coarsening.
For the recursive part we make use of the following reasonable assumptions:
\begin{itemize}
 \item A $d$-simplex that was created in a refine step will not be coarsened during the same adapt call.
 \item A $d$-simplex that was created in a coarsening step will not be refined during the same adapt call.
\end{itemize}
From these assumptions we conclude that for recursive refinement we only have to consider those $d$-simplices that
were created in a previous refinement step and that we only have to care about recursive coarsening directly after
we processed a $d$-simplex that was not refined and could be the last $d$-simplex in a family.
If refinement and coarsening are not done recursively, the runtime of \texttt{Adapt} is linear in the number
of $d$-simplices of the given forest.

An application will generally project or otherwise transform data from the
previous to the adapted mesh.
This can be done within the adaptation callback, which is known to proceed
linearly through the local elements, or after \texttt{Adapt} returns if a copy
of the old mesh has been retained.
In the latter case, one would allocate element data for the adapted mesh and then
iterate over the old and the new data simultaneously, performing the
projection in the order of the SFC.
Once this is done, the old data and the previous mesh are deallocated
\cite{BursteddeGhattasStadlerEtAl08}.

\section{Performance evaluation}

Given the design of the algorithms discussed in this paper, we expect runtimes that are
precisely proportional to the number of elements and independent of the level
of refinement.
To verify this, we present scaling and runtime tests%
\footnote{Version \texttt{v0.1} is available at
\texttt{https://bitbucket.org/cburstedde/t8code.git}}
for \texttt{New} and \texttt{Adapt} on the
JUQUEEN supercomputer at the Forschungszentrum Juelich \cite{Juqueen}, an IBM
BlueGene/Q system with 28,672 nodes consisting of
$16$ IBM PowerPC A2 @ $1.6$ GHz and $16$ GB Ram per node.
We also present one runtime study on the full MIRA system at the Argonne Leadership Computing Facility,
which has the same architecture as JUQUEEN and 49,152 nodes.
The biggest occurring number of mesh elements is around $8.5\times 10^{11}$
tetrahedra with $13$ million elements per process.

\begin{figure}
\center
\begin{minipage}{0.48\textwidth}
   \resizebox{\textwidth}{!}{\includegraphics{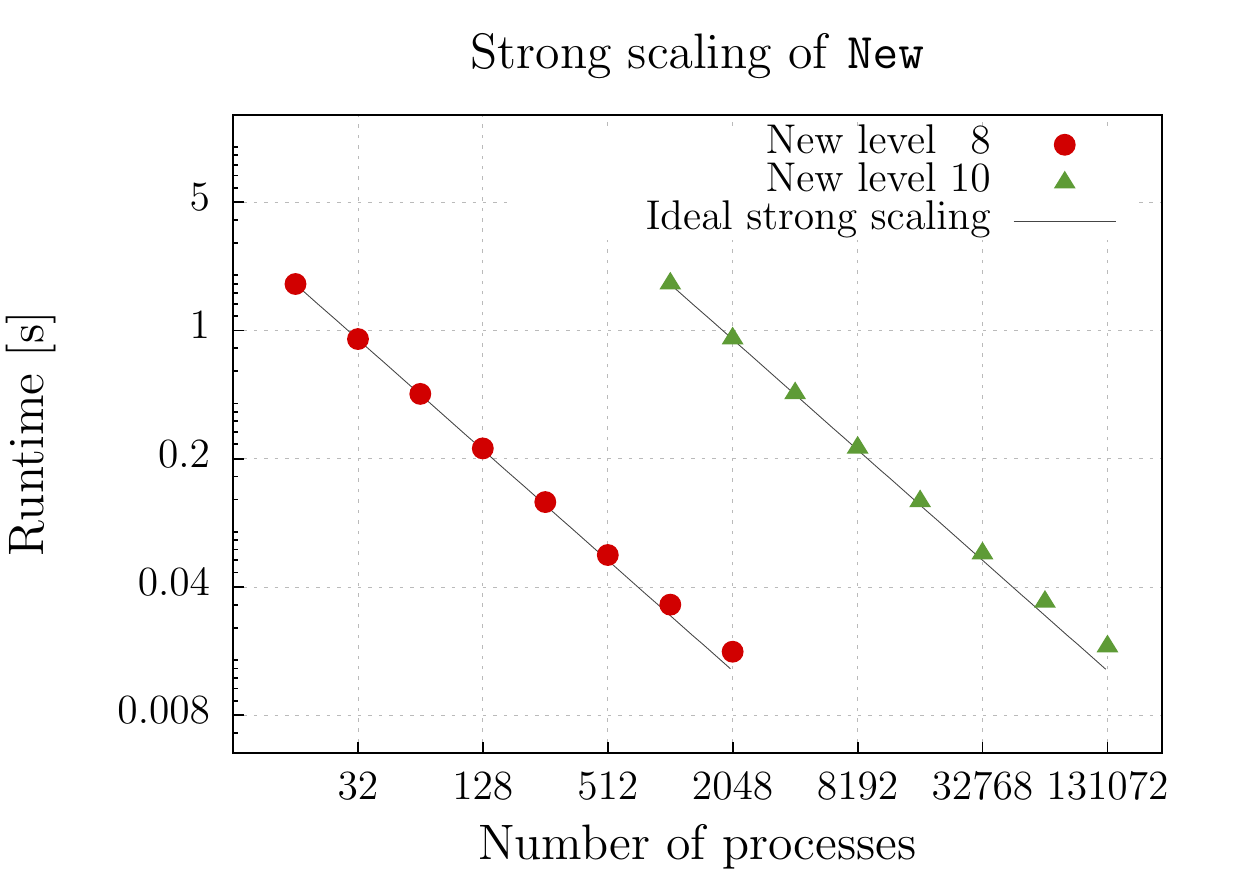}}
\end{minipage}
\begin{minipage}{0.5\textwidth}
\raisebox{18ex}{
\resizebox{0.96\textwidth}{!}{
\begin{tabular}{|lrrrr|}
 \multicolumn{5}{c}{Runtime tests for \texttt{New}}\\[2ex] \hline
 \mytabvspace 
 $\#$Procs&Level & $\#$Tetrahedra & Runtime [s] & Factor\\ \hline
 \mytabvspace
  64  & $7$ & $1.073\e9\hphantom{0}$  &  $0.059$ & --  \\
      & $8$ & $8.590\e9\hphantom{0}$  &  $0.451$ & $7.64$ \\
      & $9$ & $6.872\e10$ &  $3.58$\hphantom{1}  & $7.94$ \\
      &$10$ & $5.498\e11$ & $28.6$\hphantom{11}& $7.99$ \\ \hline
 \mytabvspace
  256 & $8$ & $8.590\e9\hphantom{0}$ & $0.116$ & --\\
 &$9$ & $6.872\e10$ & $0.898$ & $7.74$  \\
&$10$ & $5.498\e11$ & $7.15$\hphantom{1}& $7.96$ \\
 \hline
\end{tabular}
}
} 
\end{minipage}
\caption{Runtime tests for \texttt{New} on JUQUEEN.
       Left: Two strong scaling studies. A new uniform level 8 (circles) and level 10 (triangles) refinement of a coarse mesh of 512 root tetrahedra,
   carried out with 16 up to 2,048 processes and 1,024 up to 131,072 processes with 16 processes per compute node.
       Right: The data shows that the runtime of \texttt{New} is linear in the number of generated elements
          and does not additionally depend on the level.
          The uniform refinement is created from a coarse mesh of 512 root tetrahedra.
          For the first computation on 64 processes we use 1 process per compute node and
          for the computation on 256 processes we use 2 processes per node.}
\label{fig:scalenew}
\end{figure}

 The first two tests are a strong scaling (up to 131k processes) and a runtime
study of \texttt{New} in 3D, shown in Figure \ref{fig:scalenew}.  For both
tests we use a coarse mesh of $512$ tetrahedra.
 We time the \texttt{New} algorithm with input level 8 (resp.\ level 10 for higher numbers of processes).
 We execute the runtime study to examine whether \texttt{New} has the proposed level-independent linear runtime in the number of generated tetrahedra,
 which can be read from the results presented in the Table in Figure \ref{fig:scalenew}.

\begin{figure}
\center
\begin{minipage}{0.48\textwidth}
 {\includegraphics{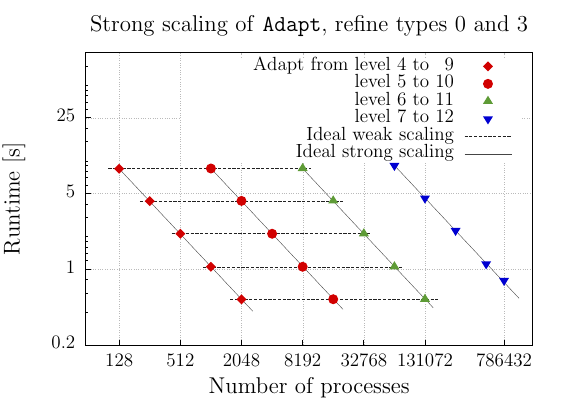}}
\end{minipage}
\hfill
\begin{minipage}{0.42\textwidth}
   \includegraphics[width=0.99\textwidth, trim=150 100 100 40, clip]{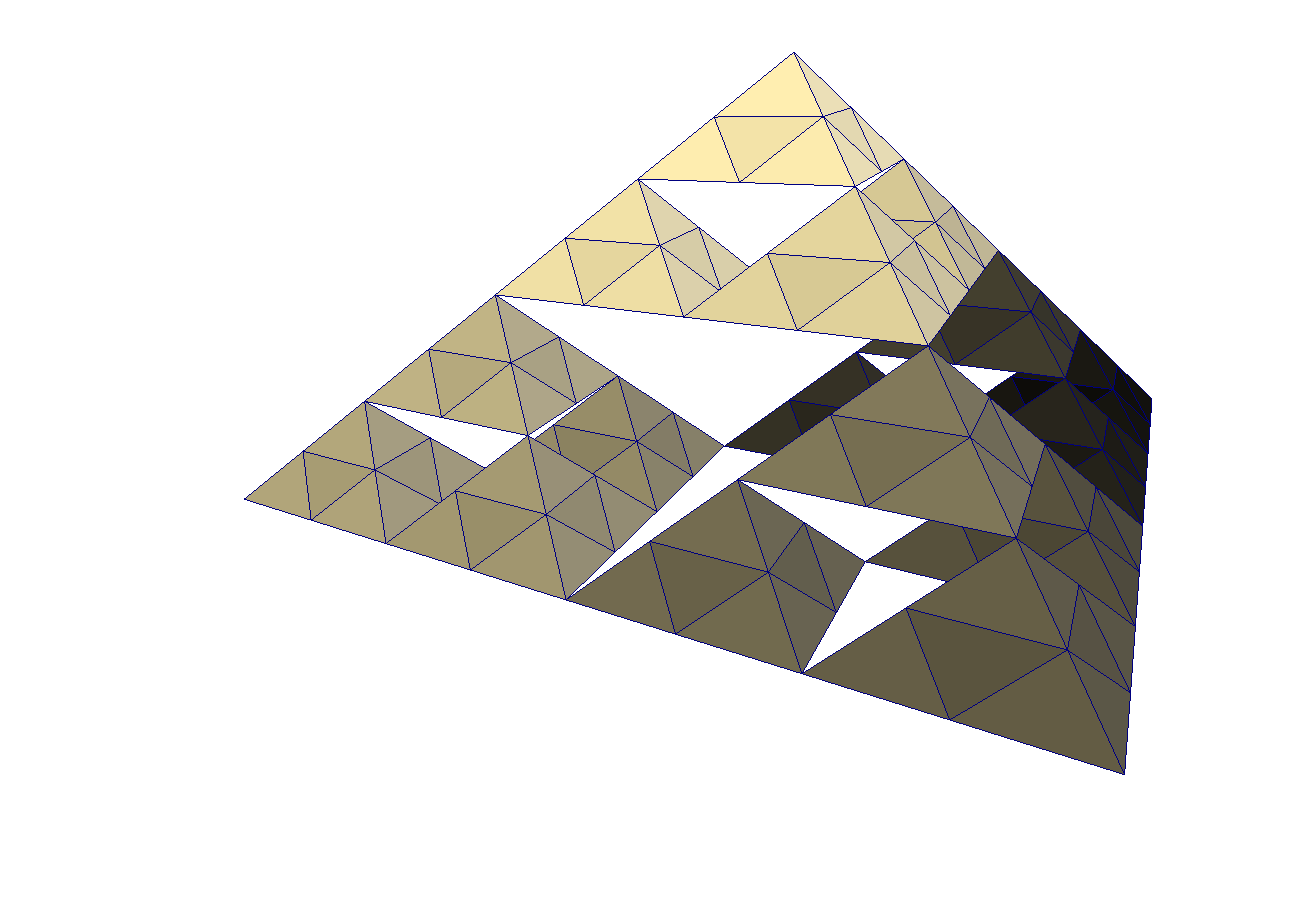}
\end{minipage}
\caption{Strong scaling for \texttt{Adapt} with a fractal refinement pattern. Starting from an initial level $k$ on
      a coarse mesh of 512 tetrahedra we refine recursively to a maximal final level $k+5$.
      The refinement callback is such that only subtetrahedra of types 0 and 3 are refined.
      Left: Strong and weak scaling on JUQUEEN with up to 131,072 processes
          and strong scaling on MIRA with up to 786,432 processes.  On both
      systems we use 16 processes per compute node.
          The level 12 mesh consists of 858,588,635,136 tetrahedra.
      Right: An initial level 0 and final level 3 refinement according to the fractal pattern.
      The subtetrahedra of levels 1 and 2 are transparent.}
\label{fig:adaptType03}
\end{figure}

 The last test is \texttt{Adapt} with a recursive nonuniform refinement pattern.
 The starting point for all runs is a mesh obtained by uniformly refining a
 coarse mesh of $512$ tetrahedra to a given initial level $k$.
 This mesh is then refined recursively
 using a single \texttt{Adapt} call, where only the tetrahedra of types $0$ and $3$ whose
 level does not exceed the fine level $k+5$ are refined recursively.
 The resulting mesh on each tetrahedron resembles a fractal pattern
 similar to the Sierpinski tetrahedron.
 We perform several strong and weak scaling runs on JUQUEEN starting with 128
 processes and scaling up to 131,072.
 The setting is 16 processes per compute node.
 We finally do another strong scaling run on the full system of the MIRA supercomputer at the
 Argonne Leadership Computing Facility with 786,432 processes and again $16$ processes per compute node.
 Figure \ref{fig:adaptType03} shows our runtime results.

\section{Conclusion}
We present a new encoding for adaptive nonconforming
triangular and tetrahedral mesh refinement based on Bey's red-refinement rule.
We identify six different types of tetrahedra (and two types of triangles)
and prescribe an ordering of the children for each of these types that differs
from Bey's original order.
By introducing an embedding of the mesh elements into a
Cartesian coordinate structure, we define a tetrahedral Morton index that can
be computed using bitwise interleaving similar to the Morton index for cubes.
This tetrahedral Morton index shares some properties with the well-known
cubical one and allows for a memory-efficient random access storage of the mesh
elements.

Exploiting the Cartesian coordinate structure, we develop several constant-time
algorithms on simplices.
These include computing the parent, the children, and the face-neighbors of a
given mesh element, as well as computing the next and previous elements according
to the SFC.

In view of providing a complete suite of parallel dynamic AMR capabilities, the
constructions and algorithms described in this paper are just the beginning.
A repartitioning algorithm following our SFC, for example, is easy to imagine,
but challenging to implement if the tree connectivity is to be partitioned
dynamically, and if global shared metadata shall be reduced from being
proportional to the number of ranks to the number of compute nodes.
The present paper provides atomic building blocks that can be used in
high-level algorithms for 2:1 balancing
\cite{IsaacBursteddeGhattas12}
and the computation of ghost elements and generalized topology iteration
\cite{IsaacBursteddeWilcoxEtAl15};
regardless, these algorithms still have to be written and are likely to raise
questions that we will have to address in future work.
Notwithstanding the above, we believe that the choices presented in this paper
are sustainable for maintaining extreme scalability in the long term.

\section*{Acknowledgments}
The authors gratefully acknowledge support by the Bonn
Hausdorff Center for Mathematics (HCM) funded by the Deutsche Forschungsgemein-
schaft (DFG). Author Holke acknowledges additional support by the Bonn Interna-
tional Graduate School for Mathematics (BIGS) as part of HCM.

We would like to thank Tobin Isaac (University of Chicago) for his
thoughts on software interfacing.

We also would like to thank the Gauss Centre for Supercomputing (GCS) for
providing computing time through the John von Neumann Institute for Computing
(NIC) on the GCS share of the supercomputer JUQUEEN \cite{Juqueen} at the
J{\"u}lich Supercomputing Centre (JSC).
GCS is the alliance of the three national supercomputing centers HLRS
(Universit\"at Stuttgart), JSC (Forschungszentrum J{\"u}lich), and LRZ
(Bayerische Akademie der Wissenschaften), funded by the German Federal Ministry
of Education and Research (BMBF) and the German State Ministries for Research
of Baden-W{\"u}rttemberg (MWK), Bayern (StMWFK), and Nordrhein-Westfalen (MIWF).
This research used resources of the Argonne Leadership Computing Facility,
which is a DOE Office of Science User Facility supported under Contract DE-AC02-06CH11357.

We are indebted to two anonymous reviewers for their careful reading of the
manuscript and their thoughtful and constructive comments.

\bibliographystyle{siamplain}
\bibliography{./ATetrahedralSFC.bbl}
\end{document}